\documentclass[sigconf]{acmart}
\usepackage{amsmath}
\usepackage{algorithm}
\usepackage{algorithmicx}
\usepackage[noend]{algpseudocode} 
\usepackage{mathtools}
\usepackage[caption=false]{subfig}
\usepackage{leftidx}
\usepackage{booktabs} 
\usepackage{todonotes}

\usepackage{amsmath,amssymb,amsfonts}
\usepackage{algorithmicx,algpseudocode}
\usepackage{subfig}
\usepackage{graphicx}
\usepackage{textcomp}
\usepackage{makecell}
\usepackage{xcolor}
\usepackage{url}
\usepackage{multirow}
\usepackage[switch]{lineno}
\usepackage{diagbox}
\usepackage{threeparttable}
\graphicspath{{./figure/} }

\def\BibTeX{{\rm B\kern-.05em{\sc i\kern-.025em b}\kern-.08emT\kern-.1667em\lower.7ex\hbox{E}\kern-.125emX}}

\renewcommand\footnotetextcopyrightpermission[1]{} 
\begin{document}

%
\title{Accelerating Sparse Approximate Matrix Multiplication on GPUs}

%

\author{Xiaoyan Liu$^1$ ,Yi Liu$^1$,Ming Dun$^2$, Bohong Yin$^1$, Hailong Yang$^1$, Zhongzhi Luan$^1$, and Depei Qian$^1$ }
\affiliation{
\institution{School of Computer Science and Engineering$^1$\\
School of Cyber Science and Technology$^2$\\ 
Beihang University$^{1,2}$, Beijing, China, 100191}
}
\email{{liuxiaoyan,yi.liu,dunming0301,yinbohong,hailong.yang,zhongzhi.luan,depeiq}@buaa.edu.cn}
%

%
\begin{abstract}

Although the matrix multiplication plays a vital role in computational linear algebra, there are few efficient solutions for matrix multiplication of the near-sparse matrices. The Sparse Approximate Matrix Multiply (SpAMM) is one of the algorithms to fill the performance gap neglected by traditional optimizations for dense/sparse matrix multiplication. However, existing SpAMM algorithms fail to exploit the performance potential of GPUs for acceleration. In this paper, we present \textit{cuSpAMM}, the first parallel SpAMM algorithm optimized for multiple GPUs. Several performance optimizations have been proposed, including algorithm re-design to adapt to the thread parallelism, blocking strategies for memory access optimization, and the acceleration with the tensor core. In addition, we scale \textit{cuSpAMM} to run on multiple GPUs with an effective load balance scheme. We evaluate \textit{cuSpAMM} on both synthesized and real-world datasets on multiple GPUs. The experiment results show that \textit{cuSpAMM} achieves significant performance speedup compared to vendor optimized \textit{cuBLAS} and \textit{cuSPARSE} libraries.

\end{abstract}

%
%
%

%

%

%
\maketitle

\section{Introduction}
\label{sec:introduction}

Generally, the existing GEMM algorithms can be classified into dense and sparse algorithms according to the ratio of non-zero elements of the input matrices. Given a matrix \textit{A} $\in \mathbb{R}^{N\times N}$, the number of non-zero elements is $O(N^{2})$ and $O(N)$ for dense and sparse algorithms, respectively. However, in real applications, there are a large number of matrices in the middle ground between dense and sparse matrices, a.k.a. near-sparse matrices, whose non-zero elements are between $O(N^{2})$ and $O(N)$. Near-sparse matrices are widely used in the field of scientific computing, such as computational chemistry~\cite{ODASHIMA2017}, quantum physics~\cite{grimme2010a}, electronic structure calculation~\cite{rudberg2018ergo}. 

The near-sparse matrices also exist in emerging domains such as deep neural networks. Especially in convolutional neural networks (CNNs), the feature and weight matrices participated in the calculation are near-sparse~\cite{cao2019seernet,gale2020sparse} due to weight pruning~\cite{ioannou2017deep} and activation functions~\cite{jaderberg2014speeding}. For example, the activation function of Rectified Linear Unit (ReLU) can lead to more than 50\% sparsity of the feature matrices on average~\cite{cao2019seernet}. In CNNs, the convolution operations between feature and weight matrices are transformed to GEMM using the \textit{im2col} algorithm~\cite{anderson2017low}. In such case, the matrices involved in the GEMM calculation are also near-sparse.

There is also a special class of matrices that are inherently near-sparse, the matrices with decay~\cite{benzi2013decay} (a.k.a. decay matrices), whose elements (values) decrease rapidly from diagonal to sides. The elements can be ignored if they are small enough and the corresponding matrices become near sparse. Due to the unique properties, there are many researches focusing on decay matrix itself, such as the decay rate~\cite{demko1984decay}, the left inverse~\cite{eijkhout1988decay,tessera2010left}, high-dimensional statistics~\cite{aune2012computation}, and numerical analysis~\cite{ye2013error}. In addition, the decay matrices often appear in widely used matrix operations such as matrix inverse~\cite{demko1984decay,benzi1999orderings}, matrix exponential~\cite{Iserles99howlarge}, Jacobi matrices~\cite{simon1982some}, and etc~\cite{benzi1999bounds}. Moreover, decay matrices are commonly adopted in application domains such as quantum chemistry~\cite{benzi2013decay,Bowler_2012} and quantum information theory~\cite{Cramer06correlations,cramer2006entanglement-area,eisert2010area,schuch2006quantum}.


However, existing research works~\cite{DBLP:conf/hpca/LinRB01,10.1145/2213977.2214056,DBLP:journals/siamsc/BulucG12} for dense and sparse GEMM are hardly efficient when applied to near-sparse matrices. On the one hand, the researches for dense GEMM focus on reducing the computation complexity. For example, Strassen's algorithm~\cite{DBLP:conf/hpca/LinRB01} and Williams' algorithm~\cite{10.1145/2213977.2214056} achieve $O(N^{2.8})$ and $O(N^{2.3727})$, respectively. Whereas, the complexity reduction is hardly useful for eliminating redundant computation of near-sparse matrices on the zero elements. On the other hand, the researches for sparse GEMM propose various storage formats such as CSR~\cite{DBLP:journals/siamsc/BulucG12} to store the sparse matrices compactly. However, the sparse formats can hardly benefit the near-sparse GEMM due to its non-sparse nature. Therefore, both the dense GEMM and the sparse GEMM have limited performance potential for near-sparse GEMM.

Fortunately, the approximation provides a good opportunity to boost the performance of near-sparse GEMM. For example, skipping the calculation of small enough elements of near-sparse matrices is a profitable way for performance acceleration. Based on such idea, Sparse Approximate Matrix Multiply (SpAMM)~\cite{spamm1} has been proposed for accelerating the decay matrix multiplication. For matrices with exponential decay, existing research~\cite{spamm4} has demonstrated the absolute error of SpAMM can be controlled reliably. 

In the meanwhile, with wide adoption in a large number of fields, GPUs have been proven with excellent speedup for matrix operations~\cite{ryoo2008optimization}. Especially with the advent of tensor core units provided by NVIDIA GPUs, mixed-precision techniques have been exploited to further accelerate matrix operations~\cite{olivaresamaya2010accelerating}. Although there are few research works optimizing SpAMM computation on CPUs~\cite{spamm2,spamm3,spamm4}, to the best of our knowledge, there is no GPU implementation available for accelerating SpAMM computation, especially exploiting the architectural features such as tensor core and scaling to multiple GPUs. This motivates our work in this paper to re-design the SpAMM algorithm for better adaption to GPU architecture and propose corresponding optimizations to achieve superior performance compared to the state-of-the-art GEMM libraries. Specifically, this paper makes the following contributions:

\begin{itemize}
\item We propose \textit{cuSpAMM}, a re-designed SpAMM algorithm tailored for GPU. Specifically, we adapt the calculation steps and the data access patterns of the SpAMM algorithm to the memory hierarchy and thread organization of GPU with increased parallelism and reduced memory accesses.

\item We propose several optimization schemes such as blocking strategies for the calculation kernels and utilization of tensor core for accelerating the calculation. In addition, we present a scaling method to extend \textit{cuSpAMM} to multiple GPUs.

\item We compare \textit{cuSpAMM} with highly optimized vendor libraries such as \textit{cuBLAS}~\cite{cublas} and \textit{cuSPARSE}~\cite{cusparse} on GPUs. In addition, we evaluate on two real datasets from electronic structure calculation (\textit{ergo}) and convolutional neural network (VGG13), to demonstrate the performance of \textit{cuSpAMM} on real-world applications.

\end{itemize}

The paper is organized as follows. In Section~\ref{sec:background}, we introduce the background of SpAMM algorithm and GPU optimizations. In Section~\ref{sec:methodology}, we present our re-designed SpAMM algorithm \textit{cuSpAMM}, and corresponding optimizations for performance acceleration on multiple GPUs. Section~\ref{sec:evaluation} compares our \textit{cuSpAMM} with the-state-of-art GEMM libraries and evaluates on two real-world datasets using \textit{cuSpAMM}. Section~\ref{sec:relatedwork} discusses the related works and Section~\ref{sec:conclusion} concludes this paper. 
\section{Background}
\label{sec:background}

\subsection{Decay matrix and SpAMM algorithm}
A matrix is defined as the decay matrix when its elements decrease following the decay rate from the diagonal to the sides. The decay rate can be exponential or algebraical, formulated as $|A[i][j]|<c \lambda^{|i-j|}$ and $|A[i][j]|<c/(|i-j|^{\lambda}+1)$ respectively, where \textit{A[i][j]} is the index of the element in matrix \textit{A}. The $|i-j|$ is the separation and can be replaced by other index-based distance function of the matrix or physical distance such as $|\vec{r_{i}}-\vec{r_{j}}|$ in non-synthetic cases~\cite{spamm4}. By mathematical definition, the decay matrix is quite dense due to few zero elements. However, under certain conditions (e.g., elements less than the threshold), a number of elements in the decay matrix can be treated as zeros, which renders the matrix as near-sparse.

SpAMM is an approximate matrix multiplication method that can be used on decay matrices. The problem solved by SpAMM can be described as $C = \alpha AB + \beta C$, where $\alpha$ and $\beta$ are parameters, and \textit{A}, \textit{B}, and \textit{C} are the matrices with exponential decay or fast algebraical decay~\cite{spamm1}. For convenience, the rest of the paper takes $\alpha=1, \beta=0$, and the square matrices $N \times N$. Besides, $\tau$ is a parameter for controlling the extent of approximation. The algorithm divides the input into quad-tree recursively, depicted in Equation~\ref{four}. Then, the algorithm performs multiplication of sub-matrices recursively. The density of sub-matrices is measured by the Frobenius norm (F-norm), depicted in Equation~\ref{norm}. The Algorithm~\ref{ori} shows the pseudo-code of SpAMM. The algorithm performs multiplication only if the product of norms from two sub-matrices is no smaller than the parameter $\tau$ (line 8 and line 13). 


\begin{equation}
\scriptsize
\label{four}
A^{t} = \\
\bigl(\begin{smallmatrix}
A_{0,0}^{t+1}& A_{0,1}^{t+1} \\
A_{1,0}^{t+1}& A_{1,1}^{t+1}
\end{smallmatrix}\bigr)
\end{equation}

\begin{equation}
\scriptsize
\label{norm}
||A^{t}||_{F} = \sqrt{\sum_{i=0}^{N-1}\sum_{j=0}^{N-1} (A_{i,j}^{t})^{2}}
\end{equation}

\begin{algorithm}[htbp]
\scriptsize
\caption{SpAMM algorithm}
\label{ori}
\begin{algorithmic}[1]
\State{\textbf{Input}: Matrices $A, B$, parameter $\tau$}
\State{\textbf{Output}: Multiplication result $C$}
\If{ lowest level }
\State{\textbf{return} $C = AB$}
\EndIf 
\For{i=0 to 1}
\For{j=0 to 1}
\If{$ ||A_{i,0}||_{F} ||B_{0,j}||_{F} \ge \tau$}
\State{$ T_{0} = SpAMM(A_{i,0},B_{0,j},\tau) $}
\Else
\State{$T_{0} = 0 $}
\EndIf
\If{$ ||A_{i,1}||_{F} ||B_{1,j}||_{F} \ge \tau$}
\State{$ T_{1} = SpAMM(A_{i,1},B_{1,j},\tau) $}
\Else
\State{$T_{1} = 0 $}
\EndIf
\State{$C_{i,j} = T_{0} + T_{1}$}
\EndFor
\EndFor
\State{\textbf{return} $C$}
\end{algorithmic}
\end{algorithm}

\subsection{GPU architecture and optimization}
\subsubsection{GPU architecture}

The CUDA~\cite{cu10} programming paradigm provides a classic definition of GPU architecture, with the thread and memory organized hierarchically.

\textbf{Thread hierarchy -} The thread are organized at four levels with coarsen granularity including \textit{thread}, \textit{warp}, \textit{block} and \textit{grid}. One \textit{warp} usually contains 32 \textit{threads}, and the \textit{threads} within a \textit{warp} are implicitly synchronized in execution. The \textit{warp} is the most basic unit for calculation execution and hardware scheduling. The \textit{block} consists of multiple \textit{threads}, and the \textit{threads} within the same \textit{block} can be synchronized. The \textit{grid} consists of multiple \textit{blocks}, and the \textit{blocks} within the \textit{grid} are executed in the SIMD fashion.

\textbf{Memory hierarchy -} The memory hierarchy can be divided into three levels, including \textit{register level}, \textit{shared memory level} and \textit{global memory level}. Each \textit{thread} has private registers, which is the fastest but also the most limited storage on GPU. The second fastest memory is the shared memory for each \textit{block}. The threads within the same \textit{block} can be synchronized through the shared memory. The \textit{global memory level} consists of global memory and texture memory that hosts the data transferred from the CPU.


\subsubsection{GPU optimization}

We briefly summarize the commonly used optimization strategies for high-performance matrix multiplication on GPU.

\textbf{Architecture targeted optimizations -} The blocking strategies~\cite{10.1007/3-540-53065-7_101} partition the matrices and performs calculations across GPU memory hierarchy. Memory prefetching strategies~\cite{lakshminarayana2014spare} utilize guiding statements for writing memory, explicitly creating buffers and calling primitives, which overlaps the data movement with computation. Register optimization strategies~\cite{ryoo2008optimization} achieve better performance by reducing the active registers and minimizing access to high-latency memory such as global memory. Other optimization approaches that avoid bank conflict and tune hyper-parameters~\cite{volkov2008benchmarking} (e.g., block size) are also useful for accelerating GEMM on GPU.

\textbf{Tensor core adoption -} The tensor core~\cite{markidis2018nvidia} introduced from Nvidia Pascal GPU has already been explored in many fields for further performance optimization such as linear algebra~\cite{haidar2018harnessing} and weather simulation~\cite{hatfield2019a} recently. In general, tensor core is a computation unit for Matrix-Multiply-Accumulate (MMA), formulated as $D_{m \times k} = A_{m \times n} \times B_{n \times k} + C_{m \times k}$, where the maximum number of matrix elements is 256. The matrix $A, B$ must be in FP16 precision, while matrix $C, D$ can be in FP16 or FP32 precision. The programming of tensor core is based on a special data structure named \textit{fragment}, which stores the computation data for the tensor core. The \textit{threads} in each \textit{wrap} operate on the \textit{fragments} to perform MMA calculation on tensor core.

\section{Methodology and Implementation}
\label{sec:methodology}

In this section, we will first give an overview of our re-designed SpAMM algorithm tailored for GPU, \textit{cuSpAMM}. Then, we introduce the design of two important kernels in \textit{cuSpAMM}, which adopts several optimization strategies as well as leverages tensor core to optimize the performance. In addition, we scale our implementation to multiple GPUs for processing larger matrices. Finally, we propose load balance and accuracy searching optimizations that further improve the performance of \textit{cuSpAMM}.

For the convenience of illustration, we use the following notations. The input of the algorithm are matrices $A,B$ $\in \mathbb{R}^{N\times N}$ and $\tau$, where $A,B$ are decay matrices, and $\tau$ is the approximation threshold. The output matrix is $C$. For optimization, we divide the input matrix into sub-matrices with size of \textit{LoNum}$\times$\textit{LoNum}. We use $BDIM=N/LoNum$ to denote the number of sub-matrices per row/column, where \textit{N} is divisible by \textit{LoNum}. The coordinates of the sub-matrix are represented by a square bracket. For example, $A[i,j]$ represents the sub-matrix with the starting index of $A[i\times LoNum][j \times LoNum]$. To avoid incomplete division, the matrices are padded with zeros to satisfy the above assumption.

\subsection{Overview of cuSpAMM}

\begin{figure}[htbp]
\centering
\includegraphics[scale=0.27]{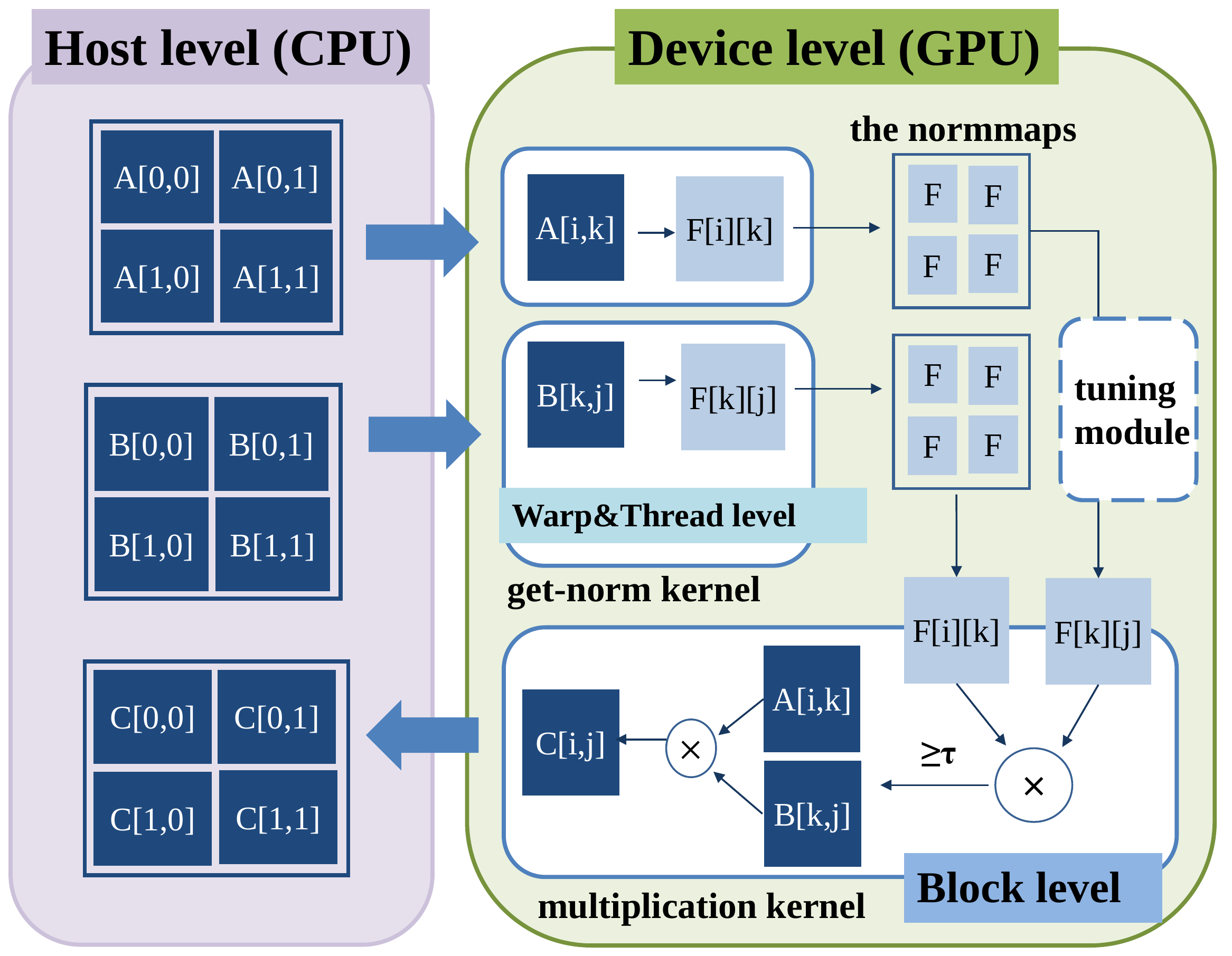}
\caption{The overview of \textit{cuSpAMM} algorithm.}
\label{overview}
\end{figure}

Figure~\ref{overview} shows the design overview of \textit{cuSpAMM}. To eliminate the GPU-unfriendly recursion in original SpAMM as well as exploit higher parallelism, we re-design the algorithm composed of two kernels, \textit{Get-norm kernel} and \textit{Multiplication kernel}. The first kernel is responsible for calculating the F-norm of input matrices, and the second kernel decides whether to multiply the matrices depending on the F-norm results from the first kernel. The array used to record the F-norm values is \textit{normmap}, where $A\_normmap[i][j] = ||A[i,j]||_{F}$ and $B\_normmap[i][j] = ||B[i,j]||_{F}$ for matrix $A, B$ respectively. The re-designed \textit{cuSpAMM} algorithm is equivalent to the original SpAMM algorithm, because they both perform calculation on the sub-matrices that satisfy the F-norm threshold ($\tau$). In addition, we propose several optimizations for the above two kernels, and utilize tensor core for further performance acceleration. Specifically, we apply the blocking optimization to \textit{cuSpAMM} across the following memory hierarchies. At \textit{device level}, we partition matrices $A, B, C$, \textit{A\_normmap} and \textit{B\_normmap} in GPU global memory. At \textit{block, warp,} and \textit{thread level}, we partition the intermediate results in corresponding memory hierarchy. The details of blocking optimization are presented in the following sections.

\subsection{\textit{Get-norm kernel}}
\begin{figure}[htbp]
\centering
\scriptsize
\subfloat[\textit{get-norm kernel}]{\includegraphics[width=1.6in, height=2.3in]{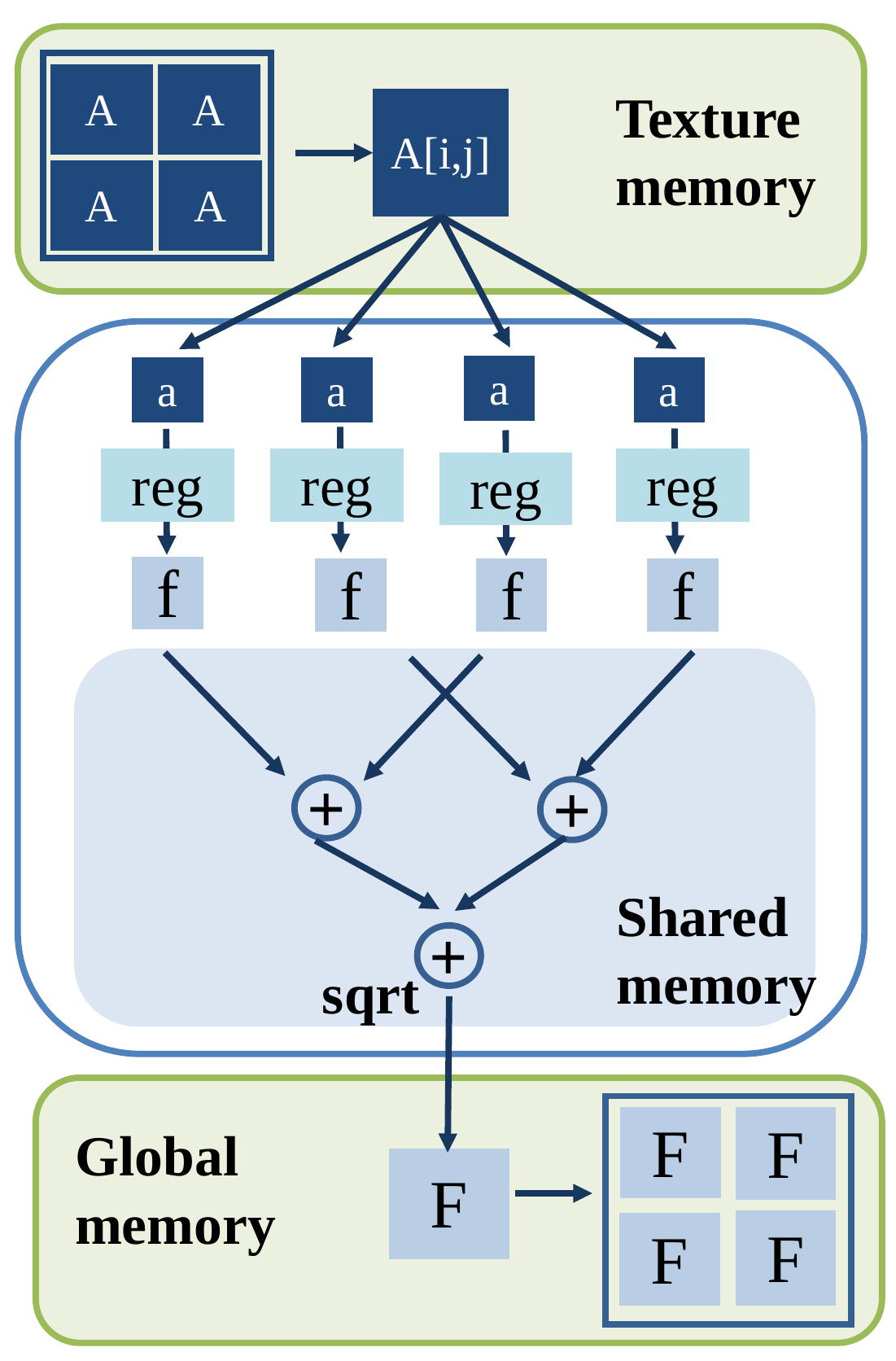}}
\subfloat[\textit{multiplication kernel}]{\includegraphics[width=1.6in, height=2.27in]{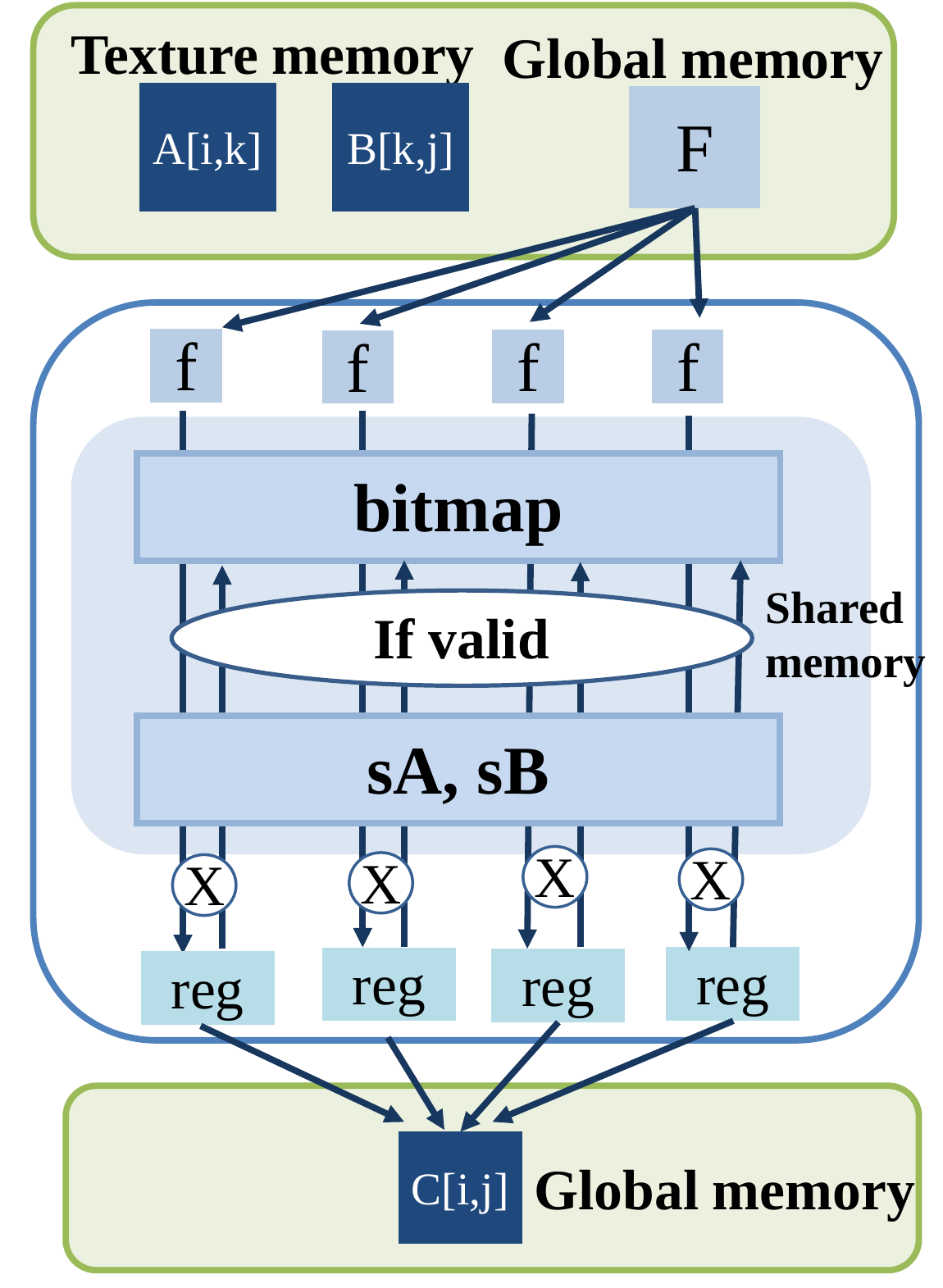}}
\caption{The blocking optimizations and execution flow of \textit{get-norm} kernel and \textit{multiplication} kernel. To illustrate, assuming LoNum=2, bank number=2, and A is the decay matrix.} \label{kernel}
\end{figure}

The \textit{get-norm} kernel is responsible for calculating the F-norm (based on Equation~\ref{norm}) results for all sub-matrices. Each block of \textit{get-norm} kernel calculates the F-norm of one sub-matrix. Considering the computation characteristics of F-norm, we adopt the reduction algorithm for better parallelization, as shown in Figure~\ref{kernel}(a). Firstly, each thread takes an element from the input matrix, calculates its square value, and stores the results into shared memory. Then, the thread block performs the reduction on the shared memory. To optimize reduction, we adopt sequential addressing instead of stride addressing to avoid the bank conflict on shared memory.

To further accelerate the performance with input matrices in FP16 precision, we use tensor core as MMA unit for reduction. Equation~\ref{redTC1} and~\ref{redTC2} show the reduction of $m^{2}$ elements, where $[1]_{m\times m}$ and $[0]_{m\times m}$ represents a square matrix composed of 1 and 0 respectively, ${x_{11}, x_{12},..., x_{mm}}$ are the data waiting for summation. After two MMA operations, the reduction results are stored in matrix \textit{D'}. This optimization can accelerate the reduction calculation compared to the traditional reduction on GPU~\cite{navarro2020gpu}. Finally, the \textit{thread 0} writes the result back to \textit{normmap}.

\begin{equation}
\scriptsize
\label{redTC1}
\begin{split}
D &=[ 1 ]_{m \times m} \times
\begin{bmatrix}
x_{11} & \cdots & x_{1m}\\
\vdots & \ddots & \vdots \\
x_{m1}& \cdots & x_{mm}
\end{bmatrix}
+ [ 0 ]_{m \times m}\\
\end{split}
\end{equation}

\begin{equation}
\scriptsize
\label{redTC2}
\begin{split}
D' &=
\begin{bmatrix}
\sum_{i=1}^{m}x_{1i} & \cdots & \sum_{i=1}^{m}x_{1i}\\
\vdots & \ddots & \vdots \\
\sum_{i=1}^{m}x_{mi}& \cdots & \sum_{i=1}^{m}x_{mi}
\end{bmatrix} \times [ 1 ]_{m \times m}
+ [ 0 ]_{m \times m} \\
\end{split}
\end{equation}

Meanwhile, we apply additional optimizations to further boost the performance. Firstly, we increase the amount of data to be processed by each thread for coalescing the global memory access. We also use the vector operations such as \textit{float2} to reduce the number of memory load instructions. Moreover, we perform loop unrolling on both algorithm level and warp level to reduces redundant jump and synchronization operations.

\subsection{\textit{Multiplication kernel}}

The \textit{multiplication} kernel is responsible for performing the actual matrix multiplication depending on the F-norm results from \textit{get-norm} kernel. Figure~\ref{kernel}(b) shows the blocking strategy and execution flow for \textit{multiplication} kernel. Each block has \textit{LoNum}$\times$\textit{LoNum} threads and is responsible for calculating one sub-matrix of matrix $C$. Supposing the block is responsible for C[i,j], and $C[i,j]=\sum A[i,k]\times B[k,j] \times bitmap[k]$, where k ranges from 0 to (BDIM-1) and bitmap[k] is 1 or 0, indicating whether A[i,k] and B[k,j] satisfies the F-norm threshold, respectively. The \textit{bitmap} is stored in shared memory. After that, all threads begin to go through the \textit{bitmap}. If \textit{bitmap[k]} is 1, the threads load the \textit{A[i,k]} and \textit{B[k,j]} into shared memory and then perform the dot product. We adopt double buffering for hiding the memory access latency during the batched sub-matrix product.

\begin{figure}[ht]
\centering
\includegraphics[scale=0.45]{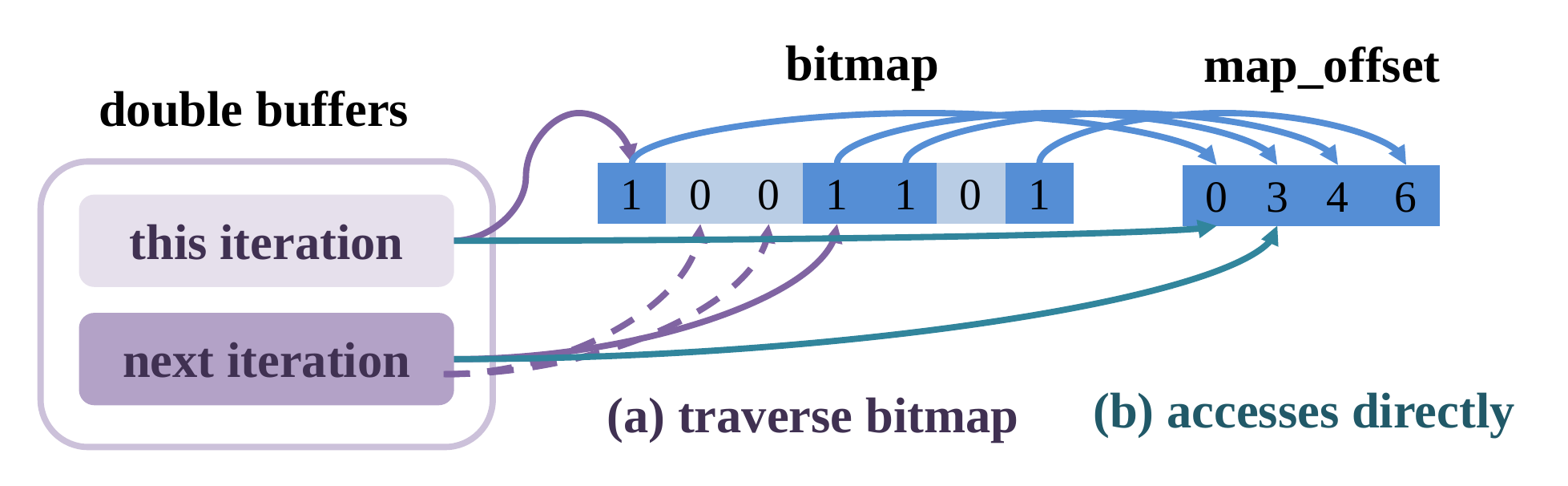}
\caption{To implement double buffering technique, (a) the naive way to traverse \textit{bitmap} introduces discontinuous accesses and additional instructions, and (b) the optimized way uses additional \textit{map\_offset} array to store the index of valid sub-matrices continuously.}
\label{buffer}
\end{figure}

However, as shown in Figure~\ref{buffer}(a), it is inefficient to implement double buffering naively. This is because the thread needs to go through the \textit{bitmap} to identify next valid sub-matrices for multiplication. The naive implementation introduces additional instructions (e.g., jump and comparison), which even leads to performance degradation. To address the above problem, we improve the double buffering technique as shown in Figure~\ref{buffer}(b), which transforms the access of the valid flags in the \textit{bitmap} from discontinuous to continuous for better locality. Although such an approach introduces additional calculations, it improves the efficiency of data prefetching with better locality, and thus accelerates the performance of \textit{multiplication} kernel.

Algorithm~\ref{mulCode} shows the optimized \textit{multipilication kernel}. The threads identify the sub-matrices that requires actual multiplication and record corresponding indexes in \textit{bitmap} in parallel (line 5$\sim$8). Specifically, threads calculate the F-norm condition using \textit{A\_normmap[i][k]} and \textit{B\_normmap[k][j]} and update \textit{bitmap[k]} for each k. We use another array \textit{map\_offset} to store the indexes of valid sub-matrices continuously (line 9$\sim$14). During each iteration (line 19$\sim$27), the first half of the block threads is responsible for matrix multiplication, and the second half is responsible for data prefetching. This strategy facilitates hiding the memory access latency by overlapping computation with data access on GPU. Besides, each thread calculates two elements of $C$, which can be stored in thread registers during dot product. 

\begin{algorithm}[htbp]
\scriptsize
\caption{\textit{cuSpAMM}: multiplication kernel}
\label{mulCode}
\begin{algorithmic}[1]
\State{\textbf{Input}: Matrix pointer *A, *B, *C, normmap pointer*A\_normmap, *B\_normmap, parameter $\tau$}
\State{\textbf{Shared memory}: bitmap, map\_offset sAW, sBW, sAR, sBR}
\For{k=threadId to N/LoNum by blockDim.x}
\State{norm\_mul = A\_normmap[i][k] $\times$ B\_normmap[k][j]}
\If{norm\_mul $\ge$ $\tau$}
\State{bitmap[i] = 1}
\Else
\State{bitmap[i] = 0}
\EndIf
\EndFor
\For{i=threadId to N/LoNum by blockDim.x}
\If{bitmap[i] == 1}
	\State{t = 0}
	\For{j=0 to i-1}
	\State{t = t + bitmap[j]}
	\EndFor
	\State{map\_offset[i] = t}
\EndIf
\EndFor
\State{\_syncthreads}
\State{reduce bitmap to get the amount of valid multiplication, save it in \textit{validNum}}
\State{if \textit{validNum} is not zero, fetch data in first block}
\State{\_syncthreads}
\For{i=0 to \textit{validNum}-1}
\State{b =  map\_offset[i]}
\State{\_syncthreads}
\State{Exchange pointer between read and write}
\If{the thread is in first half block}
\State{if i is less than \textit{validNum}-1, let next = map\_offset[i+1] and perform prefetch}
\Else
\If{the thread is in second half block}
\State{Calculate two values (c1,c2) of sAW$times$sBW}
\EndIf
\EndIf
\EndFor
\State{write back c1, c2 to matrix C}
\end{algorithmic}
\end{algorithm}

For input matrices in FP16 precision, we use the tensor core to further accelerate the matrix multiplication. Algorithm~\ref{mulTC} shows the pseudo-code of \textit{multiplication} kernel using tensor core for acceleration (with the same code in FP32 precision omitted). Each block has \textit{warpRow}$\times$\textit{warpCol} warps and each warp is responsible for the sub-matrix \textit{C[warpRow,warpCol]}. The \textit{fragment} \textit{a\_frag} and \textit{b\_frag} stores sub-matrices of \textit{A} and \textit{B}, \textit{ab\_frag} is the accumulator of intermediate results. The \textit{ab\_frag} uses FP32 precision for obtaining better accuracy. The \textit{ab\_frag} is initialized to 0. We also apply the double buffering optimization using \textit{fragment}, which is similar to the implementation in FP32 precision.

\begin{algorithm}[htbp]
\scriptsize
\caption{Multiplication kernel using tensor core}
\label{mulTC}
\begin{algorithmic}[1]
\State{\textbf{Input}: Matrix pointer *A, *B, *C, normmap pointer*A\_normmap, *B\_normmap, parameter $\tau$}
\State{\textbf{shared memory}: bitmap}
\State{\textbf{fragment}: a\_frag, b\_frag, ab\_frag}
\State{......}
\For{i=0 to \textit{validNum}-1}
\State{b =  map\_offset[i]}
\State{\_syncthreads}
\State{Exchange pointer between read and write}
\State{if i is less than \textit{validNum}-1, perform prefetch and let next = map\_offset[i+1]}
\State{load\_matrix\_sync(a\_frag, A+warpRowOff+b$\times$LoNum, N)}
\State{load\_matrix\_sync(b\_frag, B+warpColOff+b$\times$LoNum*N, N)}
\State{mma\_sync(ab\_frag, a\_frag, b\_frag, ab\_frag)}
\EndFor
\State{store\_matrix\_sync(C+warpRowOff$\times$T+warpColOff, ab\_frag)}
\end{algorithmic}
\end{algorithm}

\subsection{Scaling to multiple GPUs}

Modern servers are usually equipped with multiple GPUs (e.g., Nvidia DGX contains up to 16 GPUs). To leverage such performance potential, we extend the blocking optimizations to enable \textit{cuSpAMM} scale to multiple GPUs on a single server. Note that our multiple GPU optimizations can be further integrated with distributed matrix multiplication optimizations such as CANNON~\cite{gupta1993scalability} and SUMMA~\cite{van1997summa}. However, due to the time constraint, we focus on describing the multiple GPU optimizations on a single server, and leave the extension for distributed GPUs in future work.



Algorithm~\ref{exCode} presents the pseudo-code of scaling \textit{cuSpAMM} to multiple GPUs. Supposing that there are \textit{M} GPUs indexed from \textit{0} to \textit{M-1}. The calculation task is divided by row, and GPU \textit{i} is responsible for the rows in the range of (\textit{i}$\times$\textit{M/N}, (\textit{i}+1)$\times$\textit{M/N}] of $C$. The data transfer is divided into \textit{P} batches and implicitly managed by the use of UM~\cite{cu10} technique. We control the data transfer by ordered page faults. Firstly, several CUDA streams are created with each stream manipulating one GPU device. Then, the CPU transfers the whole matrix $B$ to each GPU in batches, and each GPU obtains the \textit{normmap} of $B$ at the same time (line 4$\sim$6). After that, the CPU sends rows [\textit{i}$\times$\textit{N/M}, (\textit{i}+1)$\times$\textit{N/M}) of matrix $A$ in batches to GPU \textit{i}. When each GPU receives the corresponding rows of $A$, it invokes \textit{get-norm} kernel and waits for the kernel to finish (line 9). After that, it invokes \textit{multiplication} kernel for calculating the result (line 11). The batching approach is able to hide the data transfer latency as well as reduce the number of active blocks, which in turn mitigates the scheduling overhead.

\begin{algorithm}[htbp]
\scriptsize
\caption{Scaling to multiple GPUs}
\label{exCode}
\begin{algorithmic}[1]
\State{\textbf{Input}: matrix A, B, $tau$}
\State{\textbf{Output}:matrix C}
\State{create CUDA stream \textit{stream} for devices}
\For{i=0 to P}
\State{launch \textit{get-norm kernel} for B}
\EndFor
\State{synchronize at stream level}
\For{i=0 to P}
\State{launch \textit{get-norm kernel} for A}
\State{synchronize at stream level}
\State{launch \textit{multiplication kernel}}
\EndFor
\State{synchronize at host level}
\State{output C}
\end{algorithmic}
\end{algorithm}

\subsection{Additional optimizations}

\subsubsection{Improving load balance} 
\begin{figure}[ht]
\centering
\includegraphics[scale=0.334]{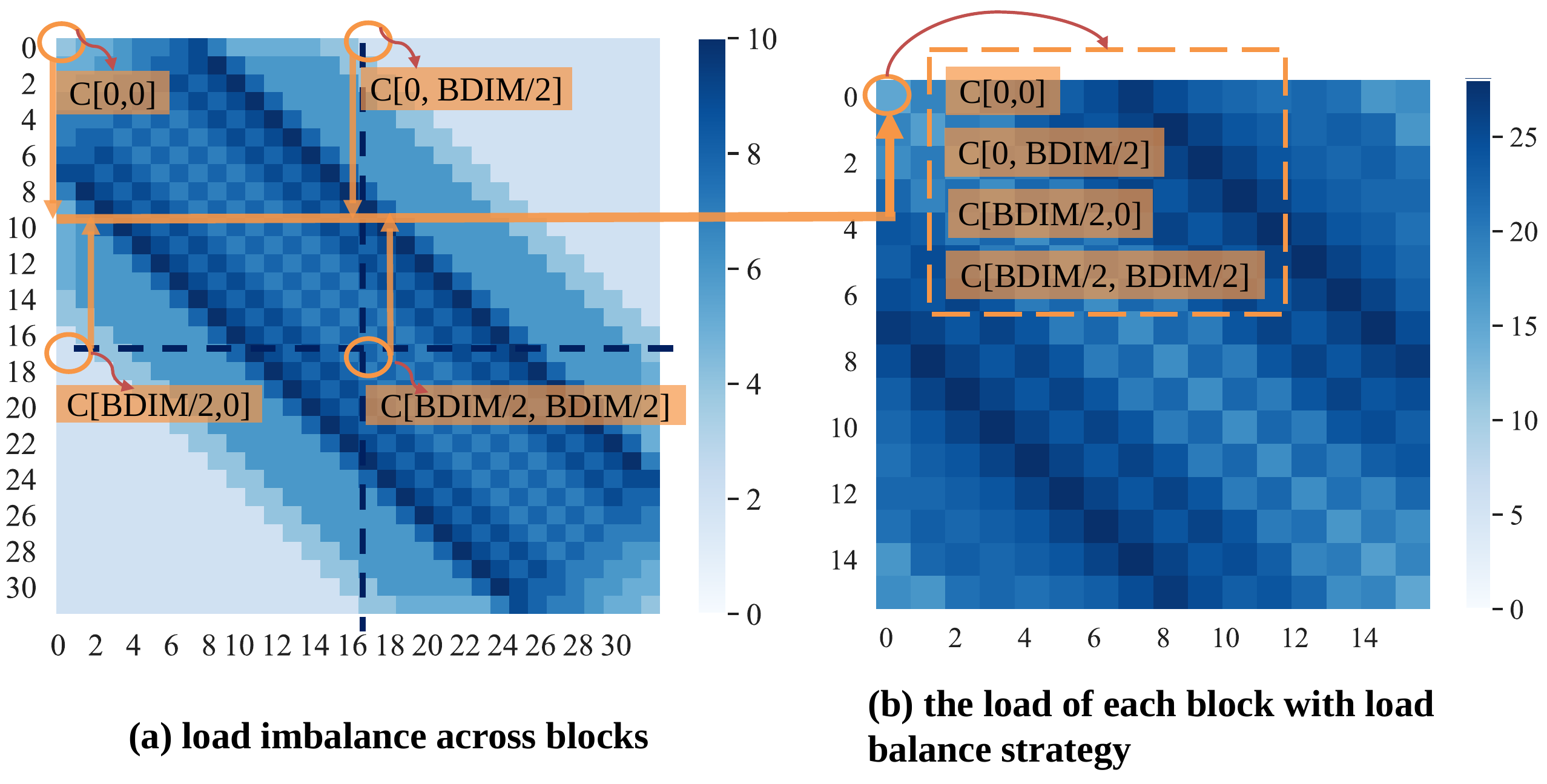}
\caption{The illustration of load balance strategy. The size of decay matrix is 1024$\times$1024, the sub-matrix is 32$\times$32 and each multiplication block is responsible for 16$\times$16 sub-matrix.}
\label{neatmap}
\end{figure}

The load imbalance could occur in \textit{multiplication} kernel as shown in Figure~\ref{neatmap}(a). This is because each block calculates the \textit{bitmap} dynamically to determine how many operations it needs to perform, which leads to block with less load staying idle and wasting resources. To measure the workload of each block, we propose the concept of valid multiplication \textit{v}. For block responsible for calculating sub-matrix $C[i,j]$, its \textit{v} equals to $\sum_{i=0}^{BDIM} bitmap[i]$. We organize the \textit{v} values of all blocks into a matrix \textit{V}, where \textit{V[i][j]} is the \textit{v} value of the sub-matrix $C[i,j]$. We observe that in matrix \textit{V}, the closer to the diagonal, the greater the \textit{v} is, which is determined by the property of decay matrix. 

Based on the above observation, we propose the following load balance strategy. Each block of the \textit{multiplication} kernel is responsible for the calculation of \textit{s} (tunable parameter) sub-matrices with equal stride. For example, as shown in Figure~\ref{neatmap}(b), one block is responsible for sub-matrices $C[0,0]$,  $C[0,BDIM/2]$, $C[BDIM/2,0]$ and $C[BDIM/2,BDIM/2]$ with \textit{s}=2. The \textit{multiplication} block can easily adopt the above strategy by adding a loop to change the index of its corresponding sub-matrices in order to achieve better load balance.

\subsubsection{Searching for customized accuracy}
\label{subsubsec:tuning}
For users using SpAMM algorithm to accelerate non-scientific applications such as deep neural networks (DNNs), adjusting $\tau$ to control the extent of approximation is not intuitive. For example, the users of DNNs are more concerned about the accuracy of the entire network, other than the numerical accuracy of a single GEMM. In such case, we provide a tuning parameter \textit{valid ratio}, formulated as \textit{valid ratio}$=\sum_{i=0,j=0}^{N-1,N-1}V[i,j]/BDIM^{3}$, to control the actual multiplication of sub-matrices, which ensures that the sub-matrices with large and dense elements participate in calculation with higher priority. This tuning parameter can better adapt to the accuracy requirements of non-scientific applications.

Specifically, after the \textit{normmaps} of $A$ and $B$ are obtained, we use a tuning kernel to calculate the average result (\textit{ave}) of the norm products of all sub-matrices. The kernel then iterates to find the suitable value of norm $\tau$ that satisfies the \textit{valid ratio} given by the user. Binary search is applied during iterations with search space [0, k$\times$\textit{ave}], where the initial value of the norm is \textit{ave}, \textit{k} is the expansion coefficient, and the upper bound of the search space is dynamically extended. The initial value of \textit{k} is one and will increase to \textit{k+1} whenever the existing upper bound cannot satisfy the search demand. Besides, users can specify the number of iterations and tolerable error of \textit{valid ratio} to balance the time cost and accuracy for searching. Since the searching algorithm is independent of the computation kernels, users can develop customized searching algorithms according to their application characteristics.


\section{Evaluation}
\label{sec:evaluation}

\subsection{Experiment setup}

\textbf{Experiment platform -} The experiments are conducted on a CPU-GPU server, with two Intel Xeon E5-2680v4 processors and eight NVIDIA Volta V100 GPUs. Each GPU contains 32 GB memory. We use CUDA v10~\cite{cu10} and nvcc compiler with -O3 option for our implementations.

\textbf{GEMM libraries -} We compare with \textit{cuBLAS} that treats the decay matrix as dense matrix, and \textit{cuSPARSE} that treats the decay matrix as sparse matrix through truncation. Note that, with truncation, the elements smaller than the threshold are treated as zero. For \textit{cuBLAS}, we use \textit{cublasSgemm} and \textit{cublasHgemm} (with tensor core optimization) for matrix multiplication in FP32 and FP16 precision respectively. For \textit{cuSPARSE}, we use \textit{cusparseScsrgemm} for matrix multiplication in FP32 precision. However, \textit{cuSPARSE} does not provide matrix multiplication in FP16 precision in CUDA v10.

\textbf{Evaluation criteria -} 
We use \textit{cudaEvent} to record the execution time of the program in seconds, and the execution time ignores the overhead of input and output transfer (including format conversion) as well as warmup time. As for accuracy criteria, we use the F-norm of error matrix $E_{n\times n}$ in Equation \ref{EF}. For \textit{cuSPARSE}, the decay matrix is truncated by setting the elements smaller than the threshold \textit{TRUN} to zeros. The \textit{nz ratio} represents the ratio of non-zero elements in the matrix. We use \textit{valid ratio} to exhibit computation and memory patterns for \textit{cuSpAMM}.

\begin{equation}  
\scriptsize
\begin{split}     
\quad E_{n\times n}=A_{n\times n}B_{n\times n}-SpAMM(A_{n\times n},B_{n\times n},\tau)
\label{EF}
\end{split}
\end{equation}

\textbf{Synthesized matrix dataset -} For performance analysis, we synthesize the matrices with algebraical decay where $a_{i,j}=b_{i,j}=0.1/(|i-j|^{0.1}+1)$, and we control the \textit{valid ratio} of the matrix indirectly by tuning the norm threshold $\tau$. Specifically, we use the tuning method in Section~\ref{subsubsec:tuning} to select the threshold and constrain the number of iterations to 20. The errors between actual and expected \textit{valid ratio} are less than 1\%.

\begin{table}[htbp]
	\renewcommand{\arraystretch}{1.3}
	\centering
	\scriptsize
	\caption{The synthesized matrices with algebraical decay.}
	\begin{tabular}{c|c|c|c|c|c|c}
		\hline
		\hline
		 \textit{valid ratio} $\backslash$ \textit{N}&1,024&2,048&4,096&8,192&16,384&32,768\\ \hline
$\approx$30\%	&1.434815&	1.310666&	1.195803&	1.093354&	0.997847&	0.905539\\ \hline
$\approx$25\%&1.456555&	1.330525&	1.222981&	1.113983&	1.012852&	0.919156\\ \hline
$\approx$20\%&	1.489164&	1.360312&	1.250158&	1.138739&	1.03536&	0.939582\\ \hline
$\approx$15\%&	1.521774&	1.40003&	1.277335&	1.171746&	1.06537&	0.966816\\ \hline
$\approx$10\%&	1.586993&	1.449676&	1.322631&	1.204753&	1.110386&	1.007668\\ \hline
$\approx$5\%&	1.695691&	1.548969&	1.413222&	1.28727&	1.170407&	1.062136\\ \hline

\end{tabular} \label{algtable}
\end{table}

\subsection{Comparison with GEMM libraries}
\label{subsec:gemmlibs}

In this section, we use synthesized matrices with algebraically decay listed in Table~\ref{algtable} for comparing with vendor optimized GEMM libraries including \textit{cuBLAS} and \textit{cuSPARSE}. 

\subsubsection{Comparison with \textit{cuBLAS}}

\begin{figure*}[htbp]
	\centering
	\scriptsize
	\subfloat[valid ratio=30\%]{\includegraphics[width=1.5in, height=1.2in]{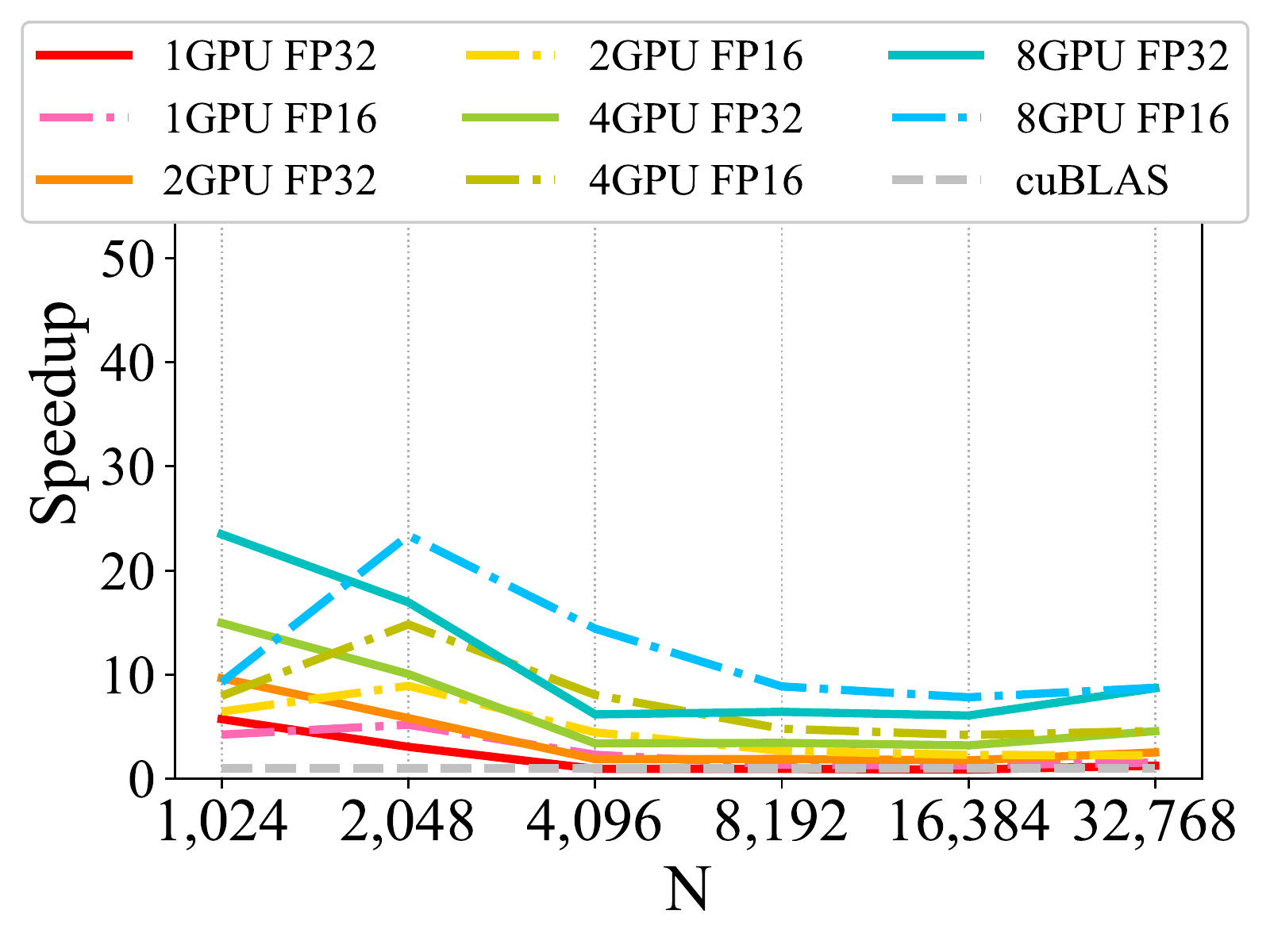}}\hspace{0.2in}
	\subfloat[valid ratio=25\%]{\includegraphics[width=1.5in, height=1.2in]{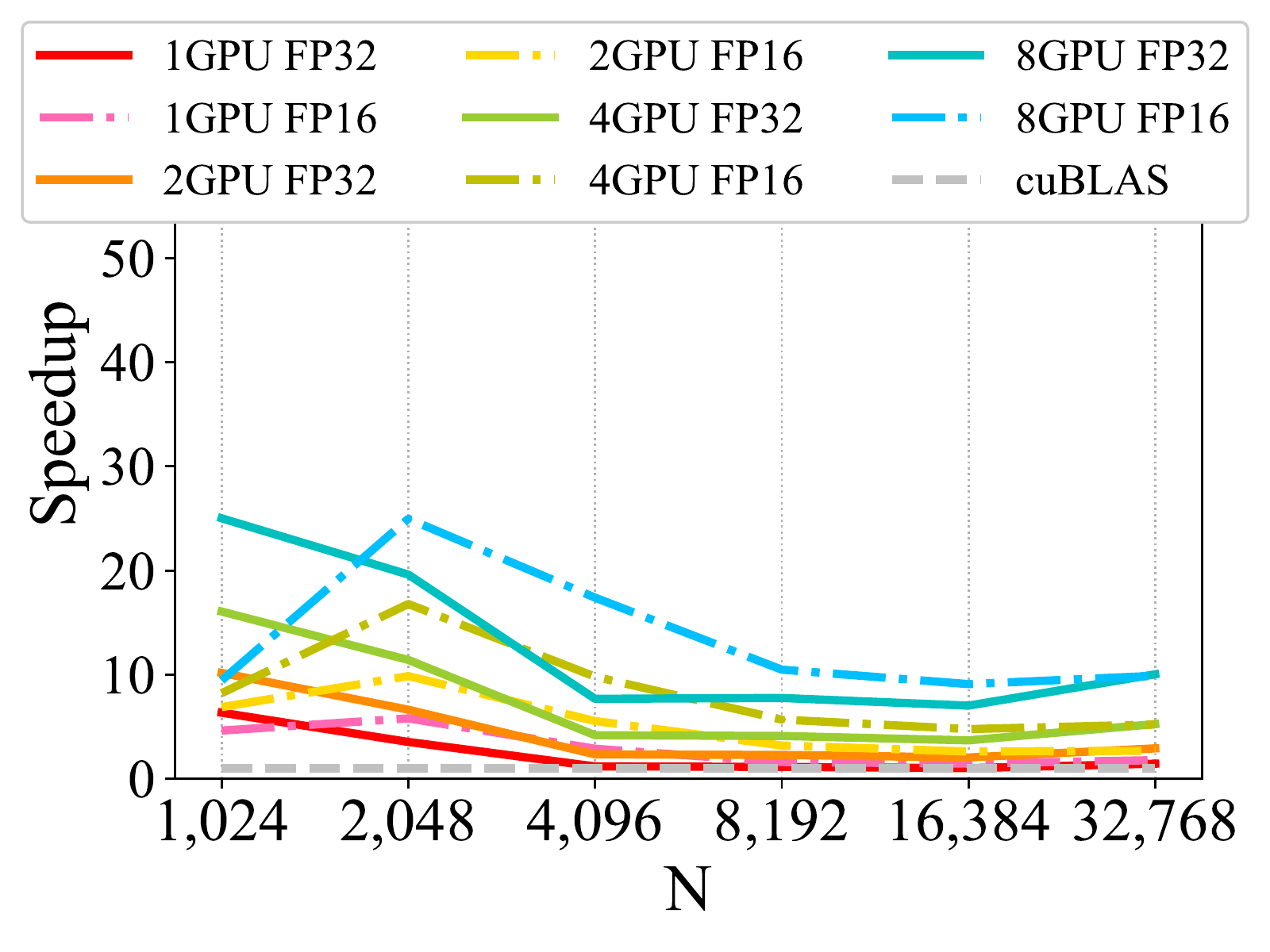}} \hspace{0.2in}
	\subfloat[valid ratio=20\%]{\includegraphics[width=1.5in, height=1.2in]{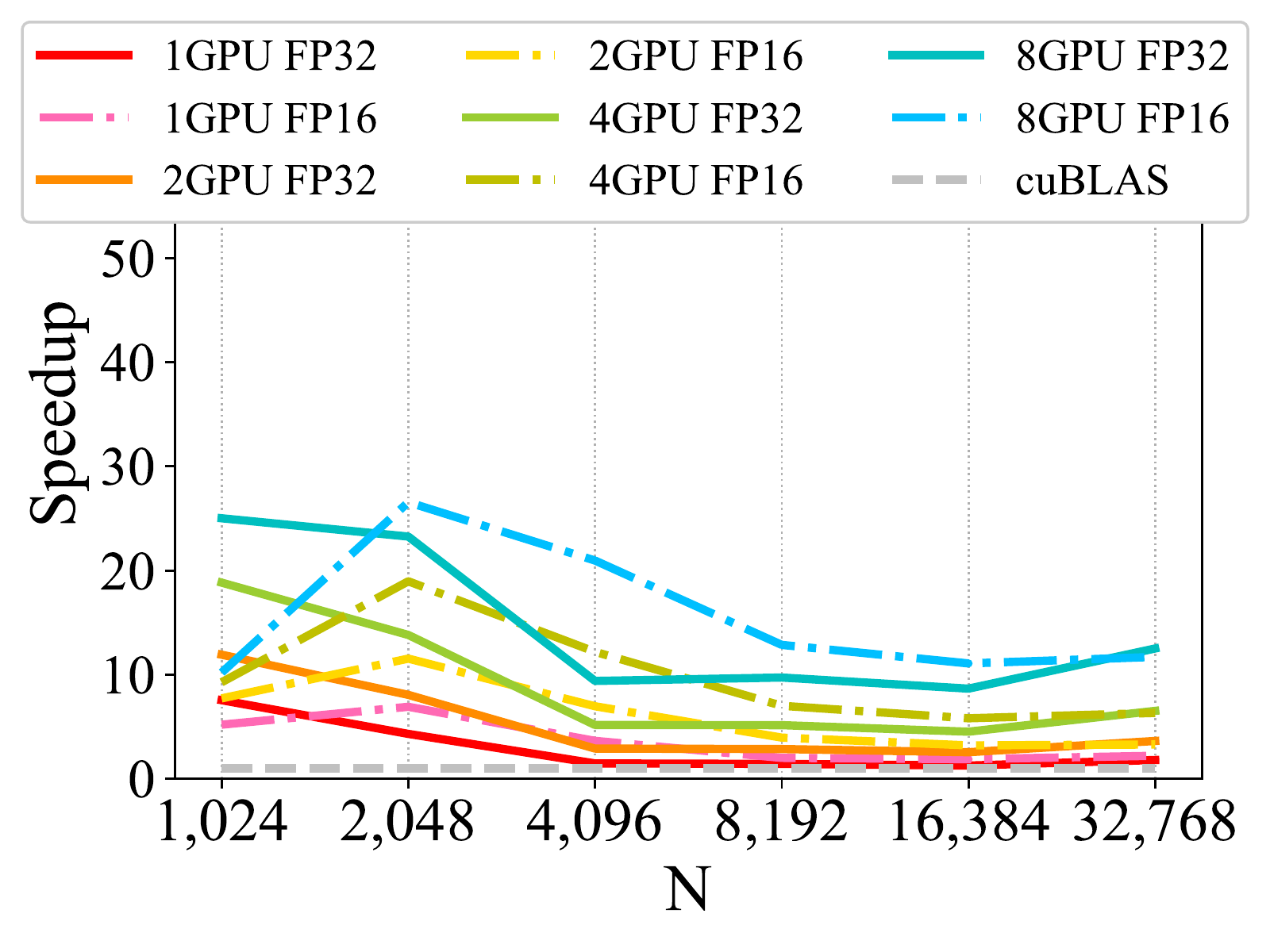}} \hspace{0.2in}
	\subfloat[valid ratio=15\%]{\includegraphics[width=1.5in, height=1.2in]{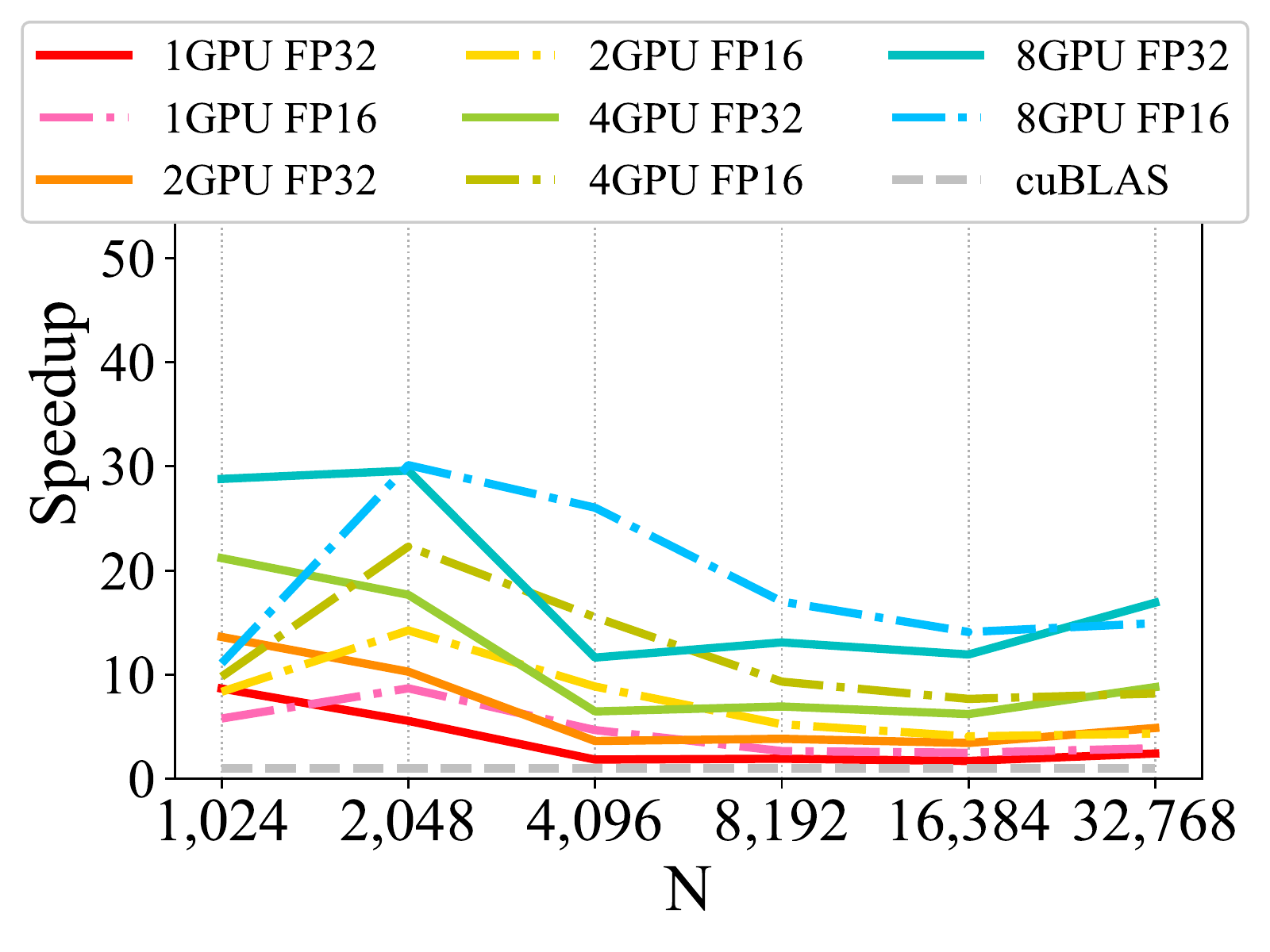}}\hspace{0.2in}
	\subfloat[valid ratio=10\%]{\includegraphics[width=1.5in, height=1.2in]{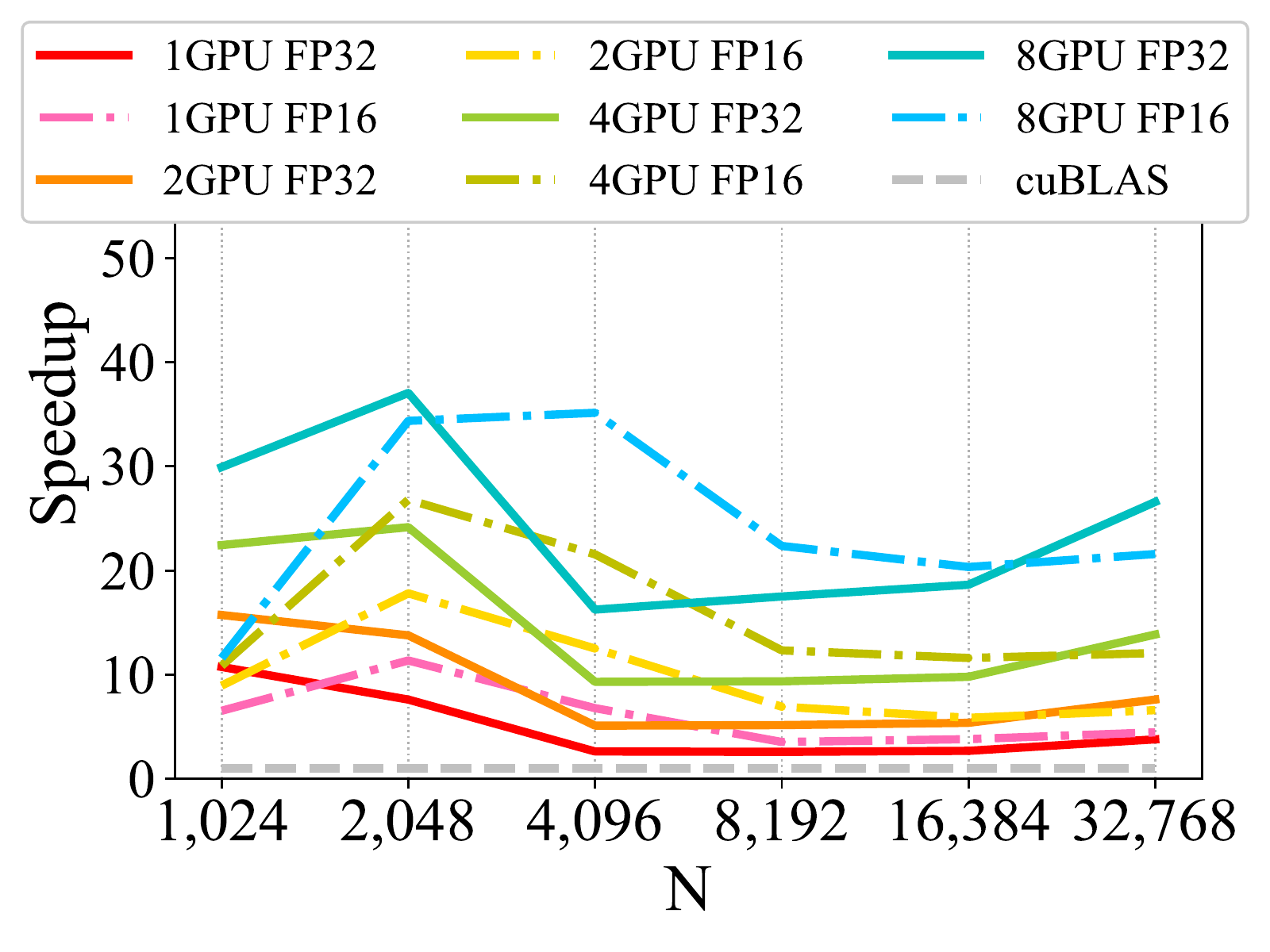}} \hspace{0.2in}
	\subfloat[valid ratio=5\%]{\includegraphics[width=1.5in, height=1.2in]{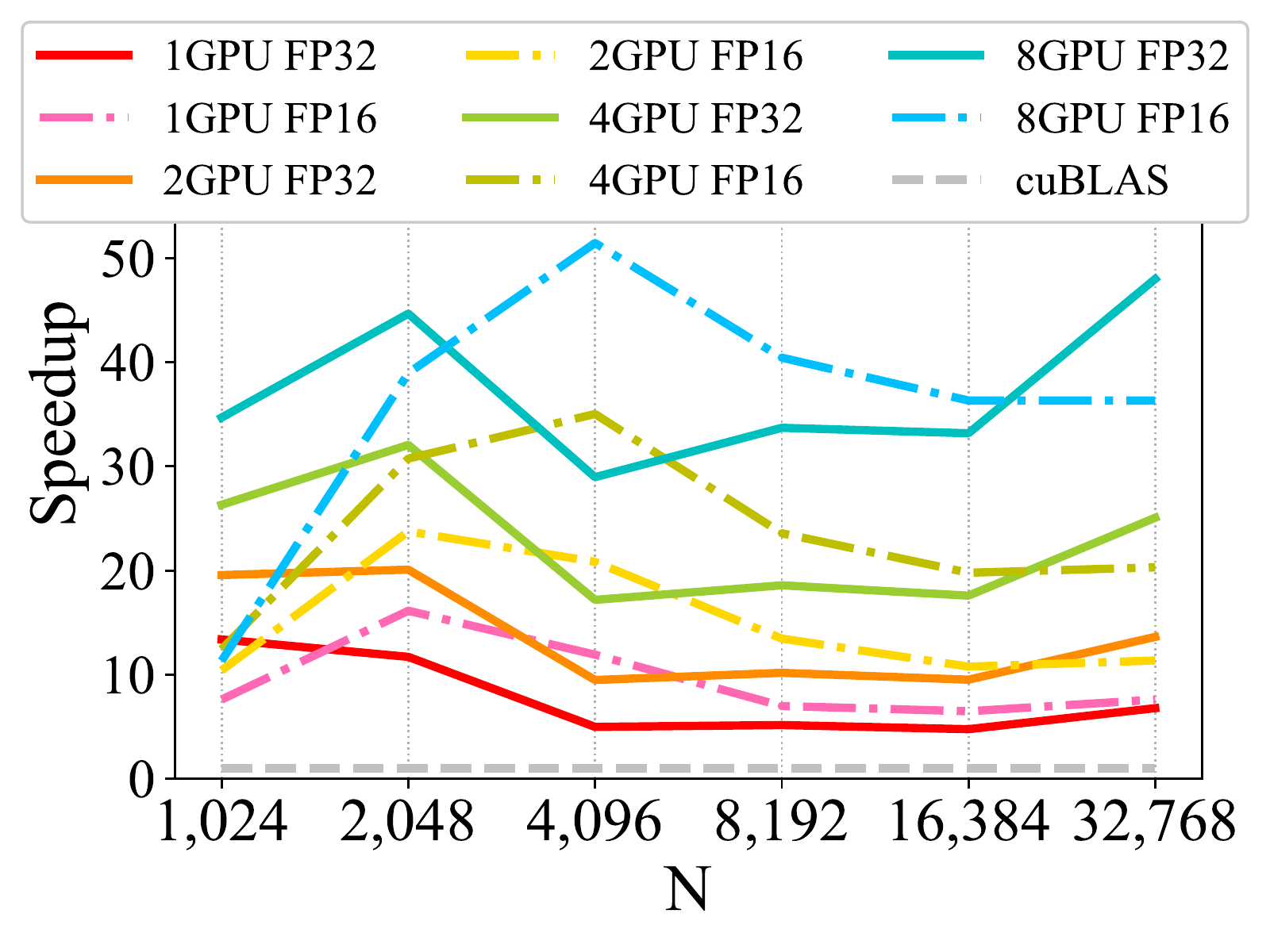}} 
	\caption{Performance comparison with \textit{cuBLAS} on matrices with algebraical decay. We change the \textit{valid ratio} from 30\% to 5\% with matrix size increasing from 1,024 to 32,768. In addition, we evaluate \textit{cuSpAMM} scaling from one to eight GPUs.} \label{cublasfig}
\end{figure*}

Table~\ref{single} presents the speedup of \textit{cuSpAMM} on a single GPU compared to \textit{cuBLAS}. The maximum speedup under each \textit{valid ratio} is highlighted in red and blue for FP32 and FP16, respectively. When the \textit{valid ratio} is 5\%, the highest speedup is achieved with 13.4$\times$ (FP32) and 16.1$\times$ (FP16). Figure~\ref{cublasfig} shows the performance comparison when scaling our optimized \textit{cuSpAMM} to multiple GPUs. It can be seen that \textit{cuSpAMM} can accelerate matrix multiplication across all matrix sizes when the \textit{valid ratio} is below a certain threshold. The reason is that when the \textit{valid ratio} is below the threshold (25\% for FP32 and 30\% for FP16), our optimizations adopted in \textit{cuSpAMM} can leverage the property of decay matrix multiplication for better parallelism compared to dense matrix multiplication adopted in \textit{cuBLAS}. Moreover, \textit{cuSpAMM} achieves better performance speedup when scaling to multiple GPUs across all matrix sizes. For example, when \textit{valid ratio} $=$ 5\%, \textit{cuSpAMM} achieves the highest speedup of 51.4$\times$ with matrix size 4,096 in FP16 running on eight GPUs, compared to \textit{cuBLAS}.

\begin{table}[htbp]
	\renewcommand{\arraystretch}{1.3}
	\centering
	\scriptsize
	\caption{The speedup on matrices with algebraical decay on a single GPU. The first row under each \textit{valid ratio} is the speedup for FP32, and the second line is for FP16.}
	\begin{tabular}{c|c|c|c|c|c|c}
		\hline
		\hline
		\textit{valid ratio} $\backslash$ \textit{N}&1,024&2,048&4,096&8,192&16,384&32,768\\ \hline
$\approx$30\%& {\textbf{\color{red}5.7  }}&3.1 &1.0 &0.9 &0.9 &1.3 \\
\cline{2-7}&4.3 & {\color{blue}\textbf{5.2}  } &2.3 &1.1 &1.3 &1.6\\ \hline
$\approx$25\%&{\color{red}\textbf{6.4}  } &3.6 &1.2 &1.2&1.0 &1.5 \\
\cline{2-7}&4.6 & {\color{blue}\textbf{5.8}  } &2.9 &1.6 &1.5 &1.8 \\ \hline
$\approx$20\%&{\color{red}\textbf{7.6}   }&4.3 &1.5 &1.4 &1.3 &1.8 \\
\cline{2-7}&5.2 & {\color{blue}\textbf{6.9}  } &3.7 &2.0 &1.9 &2.2 \\ \hline
$\approx$15\%&{\color{red}\textbf{8.7}  } &5.6 &1.9 &1.9 &1.7 &2.4 \\ 
\cline{2-7}&5.8 & {\color{blue}\textbf{8.7}  } &4.7 &2.7 &2.5 &2.9 \\ \hline
$\approx$10\%&{\color{red}\textbf{10.8}  } &7.6 &2.6 &2.6 &2.7 &3.8 \\ 
\cline{2-7}&6.5 & {\color{blue}\textbf{11.4}  }&6.8 &3.5 &3.8 &4.5 \\ \hline
$\approx$5\%&{\color{red}\textbf{13.4}  } &11.7 &5.0 &5.2 &4.8 &6.8 \\
\cline{2-7}&7.6 & {\color{blue}\textbf{16.1}  } &11.9 &7.0 &6.5 &7.6 \\ \hline
\end{tabular} \label{single}
\end{table} 

\subsubsection{Comparison with \textit{cuSPARSE}}
\label{sec:cusparse}
We choose the appropriate settings for $\tau$ so that both implementations reach the same level of error. Since determining the appropriate settings ($\tau$ and \textit{TRUN}) is time consuming, we choose matrices matrix size of 1,024 and 8,192 for illustration (larger matrix causes out-of-memory error with \textit{cuSPARSE} on a single GPU). From Table~\ref{cusparse_table}, it is clear that at the same error level, \textit{cuSpAMM} is much faster than \textit{cuSPARSE}, and the highest speedup reaches more than 601.0$\times$. In addition, the speedup becomes larger as the \textit{nz ratio} increases. Especially as the number of GPUs increases, the performance advantage of \textit{cuSpAMM} becomes even larger (e.g., 3,985.1x speedup on eight GPUs). The phenomenon further demonstrates the incapability of \textit{cuSPARSE} for handling near-sparse matrices with large non-zero ratio (e.g., more than 24\%). Note that, the execution time of \textit{cuSPARSE} does not include the time of format conversion. Thus, the performance speedup of our \textit{cuSpAMM} will be higher than the results in Table~\ref{cusparse_table} when compared in real-world application.

\begin{table}[htbp]
	\renewcommand{\arraystretch}{1.3}
	\centering
	\scriptsize
	\caption{Performance comparison with \textit{cuSPARSE}. We choose the matrix size of 1,024 (no.1) and 8,192 (no.2) with $\tau$ and \textit{TRUN} in \textit{cuSpAMM} determined for the same level of error.}
	\begin{threeparttable}
	\begin{tabular}{c|c|c|c|c|c}
		\hline
		\hline
		 \textit{no.}&\textit{nz ratio}&\textit{valid ratio}&$||E||_{F}^{\star}$&$||E||_{F}^{\dagger}$&speedup (1/2/4/8 GPUs)\\ \hline
\multirow{2}{*}{1}&52.13\%&26.83\%&1,020&996&232.3/379.5/586.8/875.8\\ 
\cline{2-6}&24.37\%&6.70\%&1,324&1,302&34.6/53.2/71.6/88.4\\ 
\cline{2-6}&10.91\%&1.87\%&1,400&1,387&11.0/14.6/22.9/21.4\\ \hline
\multirow{2}{*}{2}&59.59\%&10.35\%&38,173&37,090&589.9/1,171.4/2,127.4/3,985.1\\
\cline{2-6}&26.95\%&0.73\%&46,340&46,340&601.0/1,150.6/1,900.8/3,097.0\\
\cline{2-6}&2.12\%&0.28\%&46,340&46,340&71.2/130.9/220.8/336.9\\ \hline
	\end{tabular} 
	\begin{tablenotes}
        \footnotesize
        \item[$\star$] the error of \textit{cuSPARSE}   
        \item[$\dagger$] the error of \textit{cuSpAMM} 
      \end{tablenotes}
	\end{threeparttable}
	\label{cusparse_table}
\end{table}

\subsection{Case study}
\label{sec:casestudy}
We choose two applications widely used in scientific computing and deep neural network to further demonstrate the performance speedup. we only compare with \textit{cuBLAS} in our case study since the experiment results in Section~\ref{subsec:gemmlibs} indicate \textit{cuBLAS} achieves better performance with near-sparse matrices compared to \textit{cuSPARSE}.

\subsubsection{\textit{ergo} application}

\textit{ergo}~\cite{rudberg2018ergo} is an electronic structure computing program widely used in a range of scientific disciplines. We use \textit{ergo} and the water cluster XYZ file~\cite{ergo} to derive the decay matrices directly. The program generates four decay matrices with exponential rate, and the size of each matrix is 13,656$\times$13,656. We use \textit{cuSpAMM} to calculate the power of these matrices, and we use parameter $\tau$ to control the error ($||E_{n\times n}||_{F}$) of the results. 

\begin{figure*}[htbp]
	\centering
	\scriptsize
	\subfloat[matrix no. 1]{\includegraphics[width=1.5in, height=1.2in]{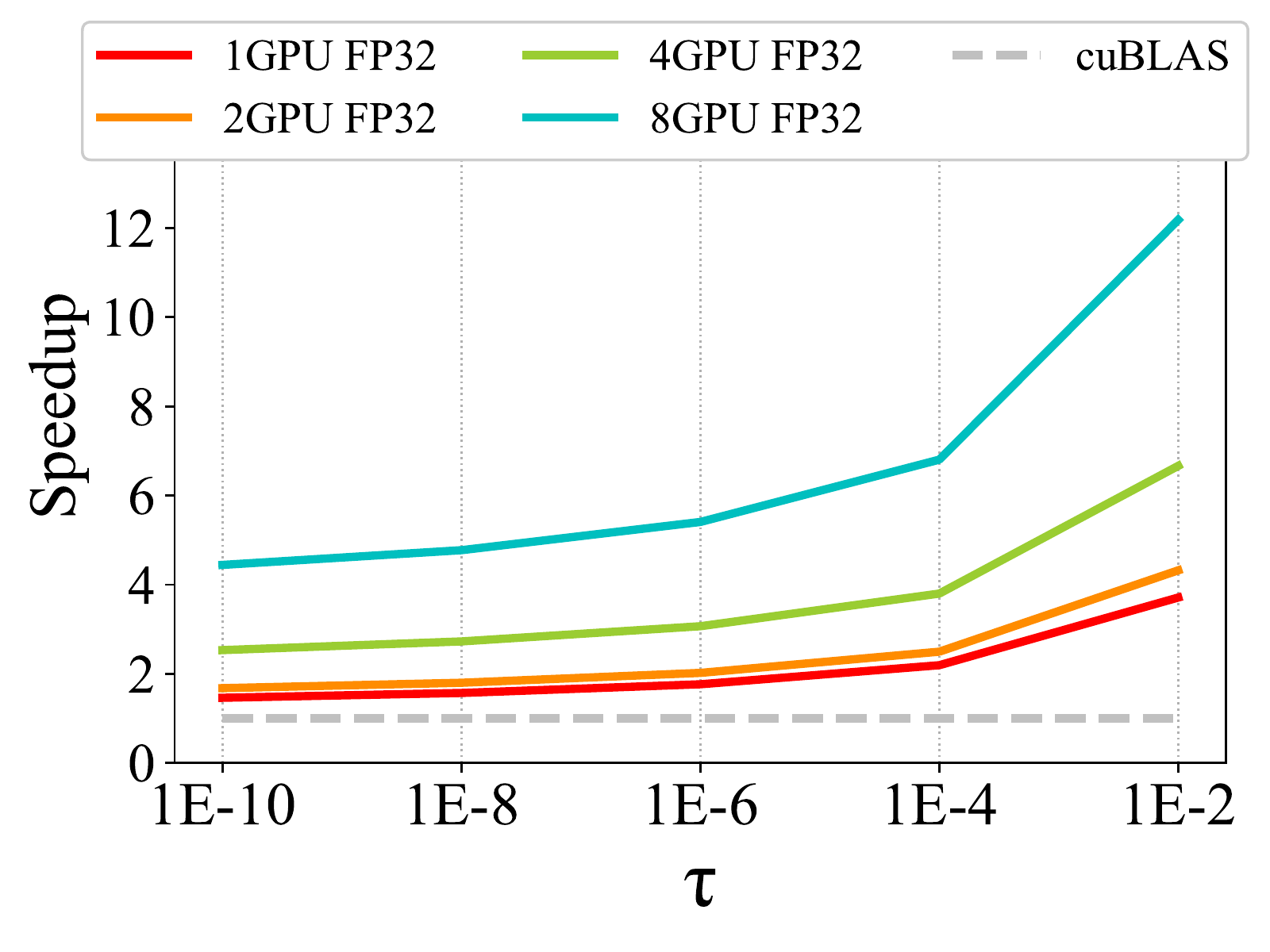}} \hspace{0.2in}
	\subfloat[matrix no. 2]{\includegraphics[width=1.5in, height=1.2in]{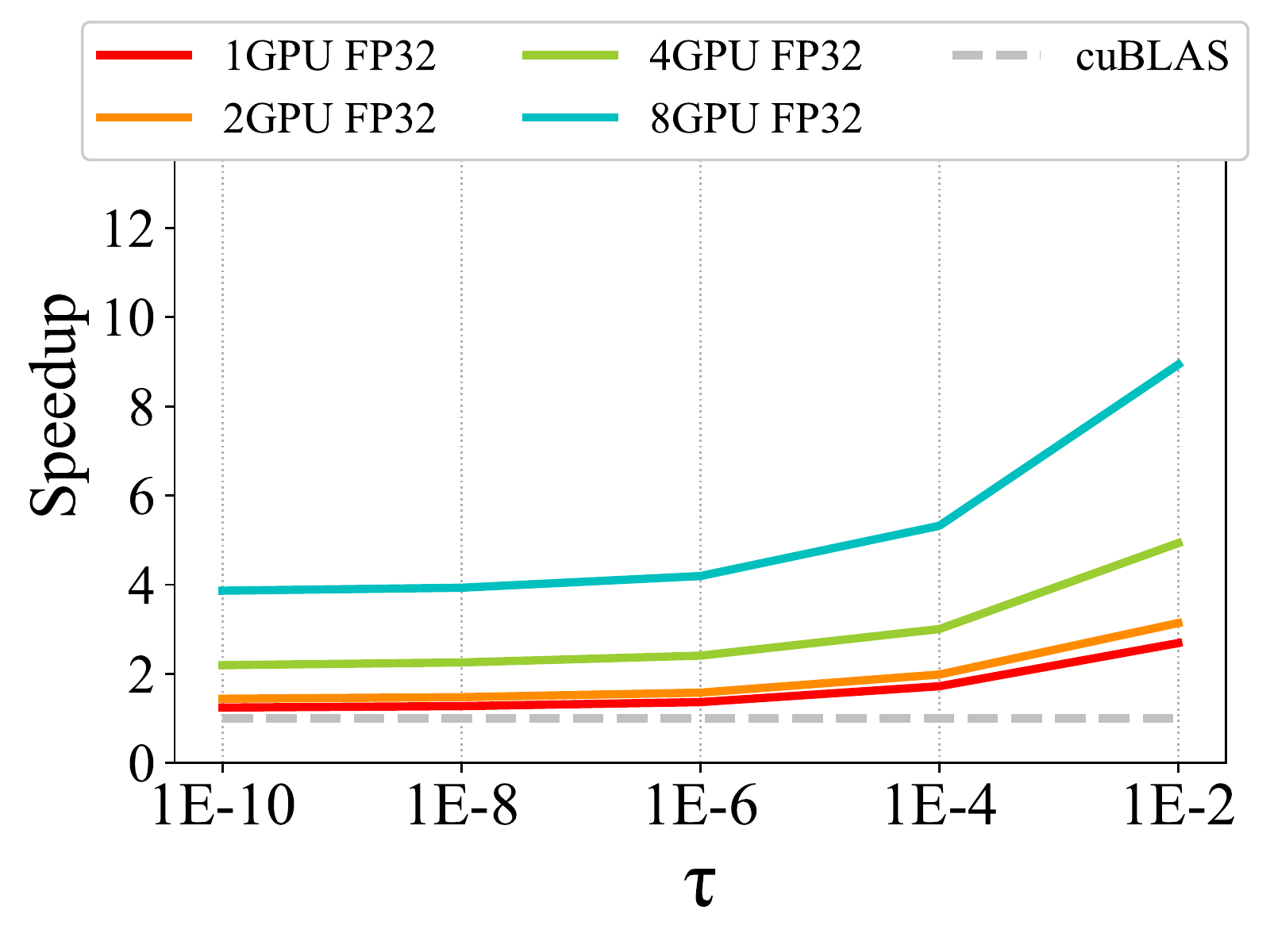}} \hspace{0.2in}
	\subfloat[matrix no. 3]{\includegraphics[width=1.5in, height=1.2in]{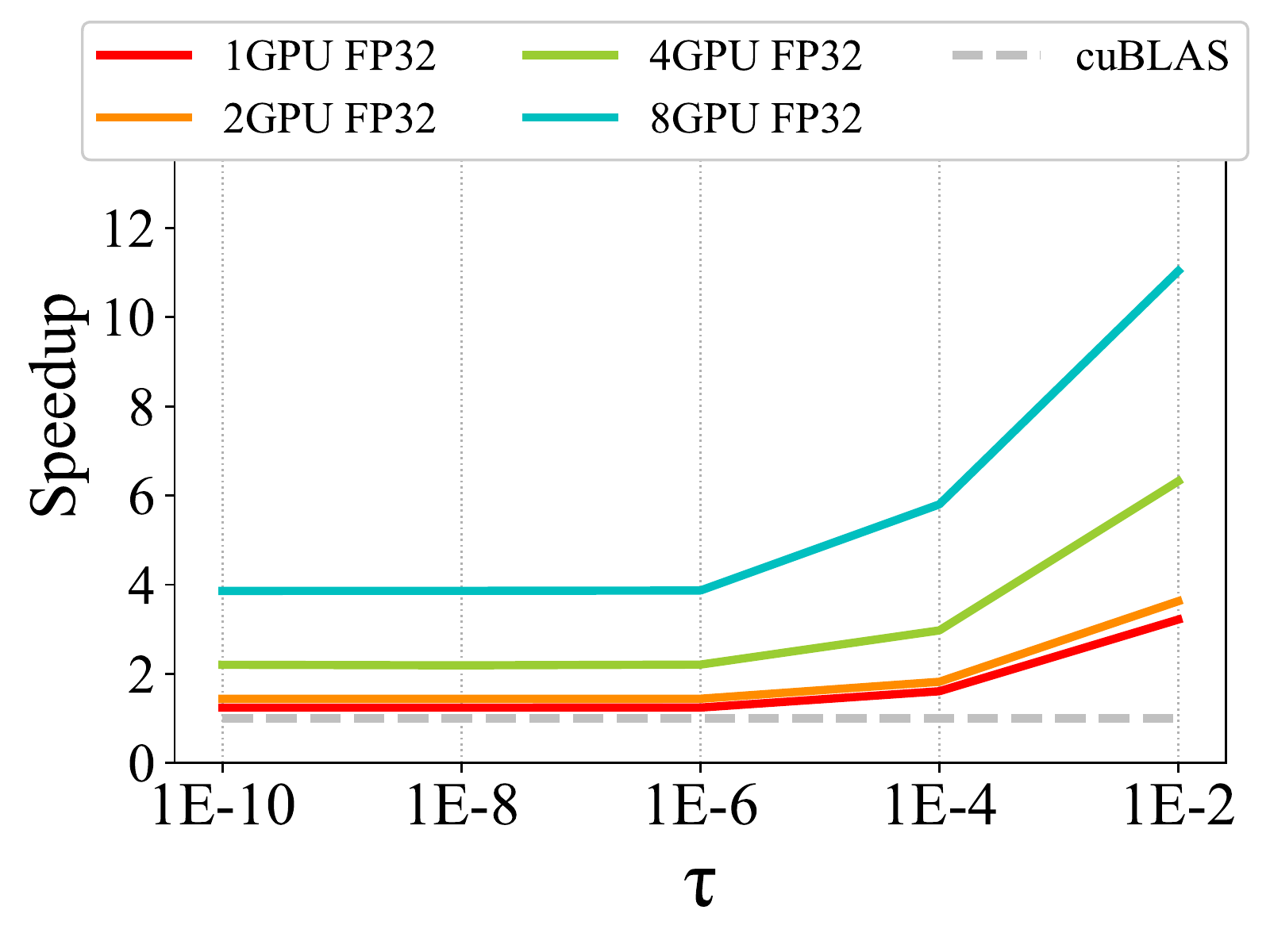}}\hspace{0.2in}
	\subfloat[matrix no. 4]{\includegraphics[width=1.5in, height=1.2in]{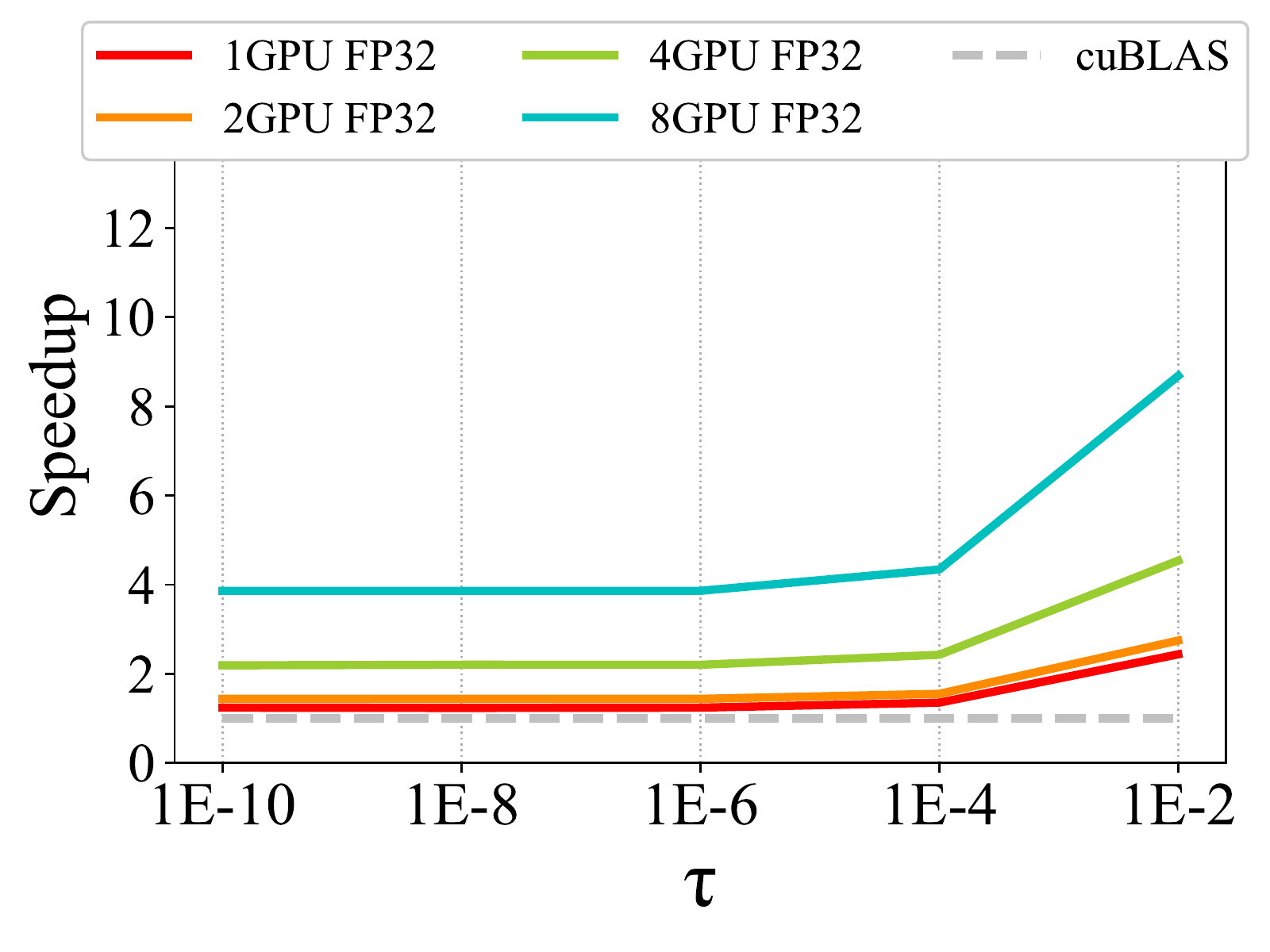}} \hspace{0.2in}
	\caption{Performance comparison with \textit{cuBLAS} on four matrices from \textit{ergo} application. The four matrices exhibit exponential decay. In addition, we evaluate \textit{cuSpAMM} scaling from one to eight GPUs.} \label{mtxfig}
\end{figure*}

Table~\ref{mtxtable} and Figure~\ref{mtxfig} present the F-norm of the matrices, the error of \textit{cuSPAMM} with different $\tau$, and the performance speedup. \textit{cuSPAMM} achieves increasing speedup when $\tau$ becomes larger across all matrices. The performance speedup of \textit{cuSPAMM} also scales when parallelizing on multiple GPUs. Especially for matrices with large F-norm ($||C||_F>1e^{7}$) and $\tau$=$1e^{-2}$, the average speedup ranges from 3.0$\times$ to 9.8$\times$ when scaling to multiple GPUs. In the meanwhile, the error introduced by \textit{cuSpAMM} is much smaller than the data involved in the calculation ($||E||_F/||C||_F <8.9e^{-7}$). For the matrix with small F-norm such as the matrix no.1 and no.2, acceptable error ($||E||_F/||C||_F<1.6e^{-5}$ when $\tau=1e^{-4}$) can be achieved with average speedup of 1.7$\times$/2.9$\times$/3.4$\times$/6.5$\times$ (1/2/4/8 GPUs). In the extreme case with no errors introduced (when $\tau=1e{-10}$), \textit{cuSpAMM} can still provide average speedup of 1.3$\times$/1.5$\times$/2.3$\times$/4.0$\times$ across all matrices.

\begin{table}[htbp]
	\renewcommand{\arraystretch}{1.3}
	\centering
	\scriptsize
	\caption{The F-norm and error of the matrices from \textit{ergo} application under different settings of $\tau$ when using \textit{cuSpAMM}.}
	\begin{tabular}{c|c|c|c|c|c|c}
		\hline
		\hline
		 \multirow{2}{*}{matrix no.}&\multirow{2}{*}{\textbf{$||C||_F$}}&\multicolumn{5}{c}{\textbf{$\tau$}} \\ 
		\cline{3-7}&&$1e^{-10}$&$1e^{-8}$&$1e^{-6}$&$1e^{-4}$&$1e^{-2}$\\ \hline
1&755&0.0 &1e-06 &9e-05 &0.01139 &1.492293  \\ \hline
2&10,406&0.0 &1e-06 &8.2e-05 &0.013414 &1.571806 \\ \hline
3&3,169,858&0.0 &0.0 &3.3e-05 &0.021516 &2.835374 \\ \hline	
4&17,171,990&0.0 &0.0 &3e-06 &0.013709 &2.102697 \\ \hline
\end{tabular} \label{mtxtable}
\end{table}

\subsubsection{VGG13 application}
We use VGG13 model on dataset MNIST. We use the \textit{im2col} algorithm to convert the trained weights and input data into matrices. We use 80\% of the dataset for training and the rest for validation. The size of the input figure is 32$\times$32, and the number of channels is three. The \textit{batch size} is set to 100 in both training and testing. The prediction accuracy of the original model is 96.6\%.


Due to the time constraint, we only choose the two largest convolution layers \textit{conv21} and \textit{conv31} from VGG13 for detailed evaluation with \textit{cuSpAMM}. Other convolution layers should expect similar tendency on performance speedup. After applying \textit{im2col} operation, the scale of the matrix multiplication is $128\times 576 \times 25,600$ and $256\times 1,152\times 6,400$ for these two layers, respectively. We apply \textit{cuSpAMM} to accelerate the two layers and use the same dataset to test the performance and accuracy. Since the size of matrices is not large enough to occupy more than four GPUs, we only perform performance evaluation on four GPUs at the largest scale for VGG13.

Table~\ref{cnntable} shows the evaluation results on both accuracy and performance when using \textit{cuSpAMM}. In general, \textit{cuSpAMM} is more effective for improving the performance of matrices multiplication in convolution neural network due to its insensitivity to matrix approximation. For both \textit{conv21} and \textit{conv31}, we can observe significant performance speedup under different settings of $\tau$ with negligible accuracy loss (less than 1.1\%). The performance speedup also scales well when increasing from one to four GPUs. Particularly for layer \textit{conv31}, \textit{cuSpAMM} achieves 2.6$\times$ to 7.8$\times$ performance speedup with the prediction accuracy unaffected. With 0.1\% accuracy loss, \textit{cuSpAMM} achieves 2.7$\times$ to 9.0$\times$ performance speedup. The highest performance speedup 12.4$\times$ is achieved when scaling to four GPUs with 1.1\% accuracy loss.

\begin{table}[htbp]
	\renewcommand{\arraystretch}{1.3}
	\centering
	\scriptsize
	\caption{The accuracy and speedup of \textit{cuSpAMM} on VGG13 application. The \textit{acc loss} measures the difference of prediction accuracy between the \textit{cuSpAMM} optimized and original models, where negative results indicate accuracy loss.}
	\begin{tabular}{c|c|c|c|c}
		\hline
		\hline
		layer&\textit{valid ratio}&\textit{acc loss}&$\tau$&\textbf{speedup (1/2/4 GPUs)}\\ \hline
\multirow{5}{*}{\textit{conv21}}&97.47\% &0 &0.1 &\textbf{2.8/5.0/8.4} \\
\cline{2-5}&96.84\% &0\% &0.05 &\textbf{2.8/5.1/8.5} \\
\cline{2-5}&85.00\% &-0.1\% &2.5 &\textbf{3.1/5.6/9.3} \\
\cline{2-5}&82.90\% &-0.1\% &3.0 &\textbf{3.2/5.7/9.4} \\ 
\cline{2-5}&63.41\% &-0.9\% &4.5 &\textbf{3.9/6.8/10.8} \\ \hline
\multirow{5}{*}{\textit{conv31}}&97.92\% &0 &2.5 &\textbf{2.6/4.8/7.6}\\ 
\cline{2-5}&94.21\% &0 &3.5 &\textbf{2.6/4.8/7.8}\\ 
\cline{2-5}&87.44\% &-0.1\% &4.5 &\textbf{2.7/5.0/8.1}\\ 
\cline{2-5}&74.36\% &-0.1\% &5.5 &\textbf{3.1/5.6/9.0}\\ 
\cline{2-5}&43.38\% &-1.1\% &7.5 &\textbf{4.8/8.1/12.4}\\ \hline
\end{tabular} \label{cnntable}
\end{table}
\section{Related Work}
\label{sec:relatedwork}
\subsection{Optimizing SpAMM}
The SpAMM was first introduced by Challacombe \textit{et al.}~\cite{spamm1}. They proved that the time complexity of the algorithm is $O(N log N)$ at worst for matrices with a sufficient decay rate. Compared to the existing approximate methods, such as truncation and rank reduction, SpAMM requires much less floating-point operations. However, Challacombe \textit{et al.}~\cite{spamm1} only presented empirical experiments on error behavior, other than a detailed analysis on numerical error. In addition, they provided limited study on the decay matrices generated by the Heaviside matrix function from electronic structure domain~\cite{grimme2010a}. Artemov \textit{et al.}~\cite{spamm4} studied the SpAMM algorithm with the matrices of exponential decay. They proved the absolute error behavior of SpAMM is $||E_{n\times n}||_{F} = O(N^(1/2) \times O(\tau^{p/2}))$, where $p<2$.

Until recently, the performance optimization of SpAMM mainly focuses on CPUs. Bock \textit{et al.}~\cite{spamm2} reformed the SpAMM computation recursively and implemented the algorithm in parallel. In addition, they used hashed (linkless) tree structure~\cite{gargantini1982an} to map the data and computation of sub-matrices. Moreover, the authors adopted hardware prefetching and optimized the performance using Intel Math Kernel Library (MKL) and AMD Core Math Library (ACML). Inspired from N-Body method~\cite{Gray2000NBodyPI}, Bock \textit{et al.}~\cite{spamm3} and Artemov \textit{et al.}~\cite{spamm4} leveraged parallel programming models such as Charm++~\cite{Kal1993CHARMAP} and Chunks/Tasks~\cite{Rubensson2017ChunksAT} to accelerate SpAMM, respectively. However, none of the existing works has optimized SpAMM on GPUs, which leaves the widely available GPU performance unexploited.

 
\subsection{High performance GEMM}
The high performance GEMM libraries on GPU have inspired our optimizations to SpAMM. \textit{cuBLAS} and \textit{cuSPARSE} are widely used high performance GEMM libraries heavily optimized by NVIDIA. Several optimization strategies have been proposed in \textit{cuBLAS} and \textit{cuSPARSE}. For example, the blocking strategies, instruction-level parallelism, parameter tuning, and tensor core acceleration have been successfully adopted for optimizing dense matrix multiplication in \textit{cuBLAS}. For sparse matrix multiplication in \textit{cuSPARSE}, various matrix storage formats such as COO, CSR and CSC are supported to optimize both sparse-sparse and sparse-dense matrix multiplication. 

Moreover, Huang \textit{et al.}~\cite{Huang2018ImplementingSA} reviewed the blocking strategies of GEMM in \textit{cuBLAS} and used tensor core to optimize the Strassen algorithm. Combined with optimizations such as software prefetching and parameter tuning, their implementation achieves 1.11$\times$ speedup on matrix multiplication compared to \textit{cuBLAS}. Mukunoki \textit{et al.}~\cite{MUKUNOKI2020112701} evaluated the parallelized linear algebra kernels with multiple data precisions on GPUs. Ryoo \textit{et al.}~\cite{ryoo2008optimization} summarized the general principles of matrix multiplication optimizations on GPU. Abdu \textit{et al.}~\cite{navarro2020gpu} used tensor core as matrix multiply-accumulate unit and proposed a chained reduction strategy. The above works have inspired the re-design and further optimization of SpAMM algorithm on GPU. Except the optimization of a single GEMM, batched GEMM~\cite{li2019coordinated} is a widely adopted technology that addresses small scale matrix multiplication and has already been integrated in vendor library such as \textit{cuBLAS}. However, batched GEMM is not applicable to SpAMM due to its time-consuming reduction operations for accumulating the final result.
\section{Conclusion}
\label{sec:conclusion}

In this paper, we propose the first SpAMM algorithm \textit{cuSpAMM}, tailored for acceleration on multiple GPUs. We re-design the SpAMM algorithm to parallelize the \textit{get-norm} kernel and \textit{multiplication} kernel. In addition, we apply several optimizations to improve the performance and accuracy of \textit{cuSpAMM}, including blocking strategy, tensor core acceleration, double buffering, load balance and parameter searching. Moreover, we scale \textit{cuSpAMM} to multiple GPUs for handling ever-increasing large near-sparse matrices. Our experiment results on both synthesized and real-world datasets show that \textit{cuSpAMM} can achieve significant speedup compared to vendor optimized \textit{cuBLAS} and \textit{cuSPARSE} libraries.

%

%
\bibliographystyle{ACM-Reference-Format}
\bibliography{spamm}


\begin{thebibliography}{52}


\ifx \showCODEN    \undefined \def \showCODEN     #1{\unskip}     \fi
\ifx \showDOI      \undefined \def \showDOI       #1{#1}\fi
\ifx \showISBNx    \undefined \def \showISBNx     #1{\unskip}     \fi
\ifx \showISBNxiii \undefined \def \showISBNxiii  #1{\unskip}     \fi
\ifx \showISSN     \undefined \def \showISSN      #1{\unskip}     \fi
\ifx \showLCCN     \undefined \def \showLCCN      #1{\unskip}     \fi
\ifx \shownote     \undefined \def \shownote      #1{#1}          \fi
\ifx \showarticletitle \undefined \def \showarticletitle #1{#1}   \fi
\ifx \showURL      \undefined \def \showURL       {\relax}        \fi
\providecommand\bibfield[2]{#2}
\providecommand\bibinfo[2]{#2}
\providecommand\natexlab[1]{#1}
\providecommand\showeprint[2][]{arXiv:#2}

\bibitem[\protect\citeauthoryear{??}{erg}{2020}]%
        {ergo}
 \bibinfo{year}{2020}\natexlab{}.
\newblock \bibinfo{title}{ErgoSCF}.
\newblock   (\bibinfo{year}{2020}).
\newblock
\showURL{%
\url{http://www.ergoscf.org/xyz/h2o.php}}


\bibitem[\protect\citeauthoryear{Anderson, Vasudevan, Keane, and
  Gregg}{Anderson et~al\mbox{.}}{2017}]%
        {anderson2017low}
\bibfield{author}{\bibinfo{person}{Andrew Anderson}, \bibinfo{person}{Aravind
  Vasudevan}, \bibinfo{person}{Cormac Keane}, {and} \bibinfo{person}{David
  Gregg}.} \bibinfo{year}{2017}\natexlab{}.
\newblock \showarticletitle{Low-memory gemm-based convolution algorithms for
  deep neural networks}.
\newblock \bibinfo{journal}{{\em arXiv preprint arXiv:1709.03395\/}}
  (\bibinfo{year}{2017}).
\newblock


\bibitem[\protect\citeauthoryear{Artemov}{Artemov}{2019}]%
        {spamm4}
\bibfield{author}{\bibinfo{person}{Anton~G. Artemov}.}
  \bibinfo{year}{2019}\natexlab{}.
\newblock \showarticletitle{Sparse approximate matrix multiplication in a fully
  recursive distributed task-based parallel framework}.
\newblock \bibinfo{journal}{{\em CoRR\/}}  \bibinfo{volume}{abs/1906.08148}
  (\bibinfo{year}{2019}).
\newblock
\showeprint[arxiv]{1906.08148}
\showURL{%
\url{http://arxiv.org/abs/1906.08148}}


\bibitem[\protect\citeauthoryear{Aune}{Aune}{2012}]%
        {aune2012computation}
\bibfield{author}{\bibinfo{person}{Erlend Aune}.}
  \bibinfo{year}{2012}\natexlab{}.
\newblock \showarticletitle{Computation and modeling for high dimensional
  Gaussian distributions}. \bibinfo{publisher}{Norges
  teknisk-naturvitenskapelige universitet, Fakultet for~…}.
\newblock


\bibitem[\protect\citeauthoryear{Benzi, Boito, and Razouk}{Benzi
  et~al\mbox{.}}{2013}]%
        {benzi2013decay}
\bibfield{author}{\bibinfo{person}{Michele Benzi}, \bibinfo{person}{Paola
  Boito}, {and} \bibinfo{person}{Nader Razouk}.}
  \bibinfo{year}{2013}\natexlab{}.
\newblock \showarticletitle{Decay Properties of Spectral Projectors with
  Applications to Electronic Structure}.
\newblock \bibinfo{journal}{{\em Siam Review\/}} \bibinfo{volume}{55},
  \bibinfo{number}{1} (\bibinfo{year}{2013}), \bibinfo{pages}{3--64}.
\newblock


\bibitem[\protect\citeauthoryear{Benzi and Golub}{Benzi and Golub}{1999}]%
        {benzi1999bounds}
\bibfield{author}{\bibinfo{person}{Michele Benzi} {and} \bibinfo{person}{Gene~H
  Golub}.} \bibinfo{year}{1999}\natexlab{}.
\newblock \showarticletitle{Bounds for the Entries of Matrix Functions with
  Applications to Preconditioning}.
\newblock \bibinfo{journal}{{\em Bit Numerical Mathematics\/}}
  \bibinfo{volume}{39}, \bibinfo{number}{3} (\bibinfo{year}{1999}),
  \bibinfo{pages}{417--438}.
\newblock


\bibitem[\protect\citeauthoryear{Benzi and Tuma}{Benzi and Tuma}{1999}]%
        {benzi1999orderings}
\bibfield{author}{\bibinfo{person}{Michele Benzi} {and}
  \bibinfo{person}{Miroslav Tuma}.} \bibinfo{year}{1999}\natexlab{}.
\newblock \showarticletitle{Orderings for Factorized Sparse Approximate Inverse
  Preconditioners}.
\newblock \bibinfo{journal}{{\em SIAM Journal on Scientific Computing\/}}
  \bibinfo{volume}{21}, \bibinfo{number}{5} (\bibinfo{year}{1999}),
  \bibinfo{pages}{1851--1868}.
\newblock


\bibitem[\protect\citeauthoryear{Bischof and Lacroute}{Bischof and
  Lacroute}{1990}]%
        {10.1007/3-540-53065-7_101}
\bibfield{author}{\bibinfo{person}{Christian~H. Bischof} {and}
  \bibinfo{person}{Philippe~G. Lacroute}.} \bibinfo{year}{1990}\natexlab{}.
\newblock \showarticletitle{An adaptive blocking strategy for matrix
  factorizations}. In \bibinfo{booktitle}{{\em CONPAR 90 --- VAPP IV}},
  \bibfield{editor}{\bibinfo{person}{Helmar Burkhart}} (Ed.).
  \bibinfo{publisher}{Springer Berlin Heidelberg}, \bibinfo{address}{Berlin,
  Heidelberg}, \bibinfo{pages}{210--221}.
\newblock


\bibitem[\protect\citeauthoryear{Bock and Challacombe}{Bock and
  Challacombe}{2012}]%
        {spamm2}
\bibfield{author}{\bibinfo{person}{Nicolas Bock} {and} \bibinfo{person}{Matt
  Challacombe}.} \bibinfo{year}{2012}\natexlab{}.
\newblock \showarticletitle{An Optimized Sparse Approximate Matrix Multiply for
  Matrices with Decay}.
\newblock \bibinfo{journal}{{\em SIAM Journal on Scientific Computing\/}}
  \bibinfo{volume}{35} (\bibinfo{date}{03} \bibinfo{year}{2012}).
\newblock
\showDOI{%
\url{https://doi.org/10.1137/120870761}}


\bibitem[\protect\citeauthoryear{Bock, Challacombe, and Kal{\'{e}}}{Bock
  et~al\mbox{.}}{2016}]%
        {spamm3}
\bibfield{author}{\bibinfo{person}{Nicolas Bock}, \bibinfo{person}{Matt
  Challacombe}, {and} \bibinfo{person}{Laxmikant~V. Kal{\'{e}}}.}
  \bibinfo{year}{2016}\natexlab{}.
\newblock \showarticletitle{Solvers for \emph{O} {(N)} Electronic Structure in
  the Strong Scaling Limit}.
\newblock \bibinfo{journal}{{\em {SIAM} J. Scientific Computing\/}}
  \bibinfo{volume}{38}, \bibinfo{number}{1} (\bibinfo{year}{2016}).
\newblock
\showDOI{%
\url{https://doi.org/10.1137/140974602}}


\bibitem[\protect\citeauthoryear{Bowler and Miyazaki}{Bowler and
  Miyazaki}{2012}]%
        {Bowler_2012}
\bibfield{author}{\bibinfo{person}{D~R Bowler} {and} \bibinfo{person}{T
  Miyazaki}.} \bibinfo{year}{2012}\natexlab{}.
\newblock \showarticletitle{{\textbackslash}mathcal$\lbrace$O$\rbrace$(N)
  methods in electronic structure calculations}.
\newblock \bibinfo{journal}{{\em Reports on Progress in Physics\/}}
  \bibinfo{volume}{75}, \bibinfo{number}{3} (\bibinfo{date}{feb}
  \bibinfo{year}{2012}), \bibinfo{pages}{036503}.
\newblock
\showDOI{%
\url{https://doi.org/10.1088/0034-4885/75/3/036503}}


\bibitem[\protect\citeauthoryear{Bulu{\c{c}} and Gilbert}{Bulu{\c{c}} and
  Gilbert}{2012}]%
        {DBLP:journals/siamsc/BulucG12}
\bibfield{author}{\bibinfo{person}{Aydin Bulu{\c{c}}} {and}
  \bibinfo{person}{John~R. Gilbert}.} \bibinfo{year}{2012}\natexlab{}.
\newblock \showarticletitle{Parallel Sparse Matrix-Matrix Multiplication and
  Indexing: Implementation and Experiments}.
\newblock \bibinfo{journal}{{\em {SIAM} J. Scientific Computing\/}}
  \bibinfo{volume}{34}, \bibinfo{number}{4} (\bibinfo{year}{2012}).
\newblock
\showDOI{%
\url{https://doi.org/10.1137/110848244}}


\bibitem[\protect\citeauthoryear{Cao, Ma, Xiao, Zhang, Liu, Zhang, Nie, and
  Yang}{Cao et~al\mbox{.}}{2019}]%
        {cao2019seernet}
\bibfield{author}{\bibinfo{person}{Shijie Cao}, \bibinfo{person}{Lingxiao Ma},
  \bibinfo{person}{Wencong Xiao}, \bibinfo{person}{Chen Zhang},
  \bibinfo{person}{Yunxin Liu}, \bibinfo{person}{Lintao Zhang},
  \bibinfo{person}{Lanshun Nie}, {and} \bibinfo{person}{Zhi Yang}.}
  \bibinfo{year}{2019}\natexlab{}.
\newblock \showarticletitle{Seernet: Predicting convolutional neural network
  feature-map sparsity through low-bit quantization}. In
  \bibinfo{booktitle}{{\em Proceedings of the IEEE Conference on Computer
  Vision and Pattern Recognition}}. \bibinfo{pages}{11216--11225}.
\newblock


\bibitem[\protect\citeauthoryear{Challacombe and Bock}{Challacombe and
  Bock}{2010}]%
        {spamm1}
\bibfield{author}{\bibinfo{person}{Matt Challacombe} {and}
  \bibinfo{person}{Nicolas Bock}.} \bibinfo{year}{2010}\natexlab{}.
\newblock \showarticletitle{Fast Multiplication of Matrices with Decay}.
\newblock \bibinfo{journal}{{\em CoRR\/}}  \bibinfo{volume}{abs/1011.3534}
  (\bibinfo{year}{2010}).
\newblock
\showeprint[arxiv]{1011.3534}
\showURL{%
\url{http://arxiv.org/abs/1011.3534}}


\bibitem[\protect\citeauthoryear{Corporation}{Corporation}{2020a}]%
        {cublas}
\bibfield{author}{\bibinfo{person}{NVIDIA Corporation}.}
  \bibinfo{year}{2020}\natexlab{a}.
\newblock \bibinfo{title}{Nvidia cuBLAS}.
\newblock   (\bibinfo{year}{2020}).
\newblock
\showURL{%
\url{https://developer.nvidia.com/cublas}}


\bibitem[\protect\citeauthoryear{Corporation}{Corporation}{2020b}]%
        {cu10}
\bibfield{author}{\bibinfo{person}{NVIDIA Corporation}.}
  \bibinfo{year}{2020}\natexlab{b}.
\newblock \bibinfo{title}{NVIDIA CUDA C programming guide}.
\newblock   (\bibinfo{year}{2020}).
\newblock


\bibitem[\protect\citeauthoryear{Corporation}{Corporation}{2020c}]%
        {cusparse}
\bibfield{author}{\bibinfo{person}{NVIDIA Corporation}.}
  \bibinfo{year}{2020}\natexlab{c}.
\newblock \bibinfo{title}{Nvidia cuSPARSE}.
\newblock   (\bibinfo{year}{2020}).
\newblock
\showURL{%
\url{https://docs.nvidia.com/cuda/cusparse}}


\bibitem[\protect\citeauthoryear{Cramer and Eisert}{Cramer and Eisert}{2006}]%
        {Cramer06correlations}
\bibfield{author}{\bibinfo{person}{M. Cramer} {and} \bibinfo{person}{J.
  Eisert}.} \bibinfo{year}{2006}\natexlab{}.
\newblock \showarticletitle{Correlations, spectral gap, and entanglement in
  harmonic quantum systems on generic lattices}.
\newblock \bibinfo{journal}{{\em New Journ. Phys\/}} (\bibinfo{year}{2006}).
\newblock


\bibitem[\protect\citeauthoryear{Cramer, Eisert, Plenio, and Dreissig}{Cramer
  et~al\mbox{.}}{2006}]%
        {cramer2006entanglement-area}
\bibfield{author}{\bibinfo{person}{M Cramer}, \bibinfo{person}{Jens Eisert},
  \bibinfo{person}{Martin~B Plenio}, {and} \bibinfo{person}{J Dreissig}.}
  \bibinfo{year}{2006}\natexlab{}.
\newblock \showarticletitle{Entanglement-area law for general bosonic harmonic
  lattice systems}.
\newblock \bibinfo{journal}{{\em Physical Review A\/}} \bibinfo{volume}{73},
  \bibinfo{number}{1} (\bibinfo{year}{2006}), \bibinfo{pages}{012309}.
\newblock


\bibitem[\protect\citeauthoryear{Demko, Moss, and Smith}{Demko
  et~al\mbox{.}}{1984}]%
        {demko1984decay}
\bibfield{author}{\bibinfo{person}{Stephen Demko}, \bibinfo{person}{William~F
  Moss}, {and} \bibinfo{person}{Philip~W Smith}.}
  \bibinfo{year}{1984}\natexlab{}.
\newblock \showarticletitle{Decay rates for inverses of band matrices}.
\newblock \bibinfo{journal}{{\it Math. Comp.}} \bibinfo{volume}{43},
  \bibinfo{number}{168} (\bibinfo{year}{1984}), \bibinfo{pages}{491--499}.
\newblock


\bibitem[\protect\citeauthoryear{Eijkhout and Polman}{Eijkhout and
  Polman}{1988}]%
        {eijkhout1988decay}
\bibfield{author}{\bibinfo{person}{Victor Eijkhout} {and} \bibinfo{person}{Ben
  Polman}.} \bibinfo{year}{1988}\natexlab{}.
\newblock \showarticletitle{Decay rates of inverses of banded M-matrices that
  are near to Toeplitz matrices}.
\newblock \bibinfo{journal}{{\it Linear Algebra Appl.}}  \bibinfo{volume}{109}
  (\bibinfo{year}{1988}), \bibinfo{pages}{247--277}.
\newblock


\bibitem[\protect\citeauthoryear{Eisert, Cramer, and Plenio}{Eisert
  et~al\mbox{.}}{2010}]%
        {eisert2010area}
\bibfield{author}{\bibinfo{person}{Jens Eisert}, \bibinfo{person}{M Cramer},
  {and} \bibinfo{person}{Martin~B Plenio}.} \bibinfo{year}{2010}\natexlab{}.
\newblock \showarticletitle{Area laws for the entanglement entropy - a review}.
\newblock \bibinfo{journal}{{\em Reviews of Modern Physics\/}}
  \bibinfo{volume}{82}, \bibinfo{number}{1} (\bibinfo{year}{2010}),
  \bibinfo{pages}{277--306}.
\newblock


\bibitem[\protect\citeauthoryear{Gale, Zaharia, Young, and Elsen}{Gale
  et~al\mbox{.}}{2020}]%
        {gale2020sparse}
\bibfield{author}{\bibinfo{person}{Trevor Gale}, \bibinfo{person}{Matei
  Zaharia}, \bibinfo{person}{Cliff Young}, {and} \bibinfo{person}{Erich
  Elsen}.} \bibinfo{year}{2020}\natexlab{}.
\newblock \showarticletitle{Sparse GPU kernels for deep learning}.
\newblock \bibinfo{journal}{{\em arXiv preprint arXiv:2006.10901\/}}
  (\bibinfo{year}{2020}).
\newblock


\bibitem[\protect\citeauthoryear{Gargantini}{Gargantini}{1982}]%
        {gargantini1982an}
\bibfield{author}{\bibinfo{person}{Irene Gargantini}.}
  \bibinfo{year}{1982}\natexlab{}.
\newblock \showarticletitle{An effective way to represent quadtrees}.
\newblock \bibinfo{journal}{{\em Communications of The ACM\/}}
  \bibinfo{volume}{25}, \bibinfo{number}{12} (\bibinfo{year}{1982}),
  \bibinfo{pages}{905--910}.
\newblock


\bibitem[\protect\citeauthoryear{Gray and Moore}{Gray and Moore}{2000}]%
        {Gray2000NBodyPI}
\bibfield{author}{\bibinfo{person}{Alexander~G. Gray} {and}
  \bibinfo{person}{Andrew~W. Moore}.} \bibinfo{year}{2000}\natexlab{}.
\newblock \showarticletitle{'N-Body' Problems in Statistical Learning}. In
  \bibinfo{booktitle}{{\em NIPS}}.
\newblock


\bibitem[\protect\citeauthoryear{Grimme, Antony, Ehrlich, and Krieg}{Grimme
  et~al\mbox{.}}{2010}]%
        {grimme2010a}
\bibfield{author}{\bibinfo{person}{Stefan Grimme}, \bibinfo{person}{Jens
  Antony}, \bibinfo{person}{Stephan Ehrlich}, {and} \bibinfo{person}{Helge
  Krieg}.} \bibinfo{year}{2010}\natexlab{}.
\newblock \showarticletitle{A consistent and accurate ab initio parametrization
  of density functional dispersion correction (DFT-D) for the 94 elements
  H-Pu}.
\newblock \bibinfo{journal}{{\em Journal of Chemical Physics\/}}
  \bibinfo{volume}{132}, \bibinfo{number}{15} (\bibinfo{year}{2010}),
  \bibinfo{pages}{154104}.
\newblock


\bibitem[\protect\citeauthoryear{Gupta and Kumar}{Gupta and Kumar}{1993}]%
        {gupta1993scalability}
\bibfield{author}{\bibinfo{person}{Anshul Gupta} {and} \bibinfo{person}{Vipin
  Kumar}.} \bibinfo{year}{1993}\natexlab{}.
\newblock \showarticletitle{Scalability of parallel algorithms for matrix
  multiplication}. In \bibinfo{booktitle}{{\em 1993 International Conference on
  Parallel Processing-ICPP'93}}, Vol.~\bibinfo{volume}{3}. IEEE,
  \bibinfo{pages}{115--123}.
\newblock


\bibitem[\protect\citeauthoryear{Haidar, Tomov, Dongarra, and Higham}{Haidar
  et~al\mbox{.}}{2018}]%
        {haidar2018harnessing}
\bibfield{author}{\bibinfo{person}{Azzam Haidar}, \bibinfo{person}{Stanimire
  Tomov}, \bibinfo{person}{Jack Dongarra}, {and} \bibinfo{person}{Nicholas~J
  Higham}.} \bibinfo{year}{2018}\natexlab{}.
\newblock \showarticletitle{Harnessing GPU tensor cores for fast FP16
  arithmetic to speed up mixed-precision iterative refinement solvers}. In
  \bibinfo{booktitle}{{\em SC18: International Conference for High Performance
  Computing, Networking, Storage and Analysis}}. IEEE,
  \bibinfo{pages}{603--613}.
\newblock


\bibitem[\protect\citeauthoryear{Hatfield, Chantry, D{\"u}ben, and
  Palmer}{Hatfield et~al\mbox{.}}{2019}]%
        {hatfield2019a}
\bibfield{author}{\bibinfo{person}{Sam Hatfield}, \bibinfo{person}{Matthew
  Chantry}, \bibinfo{person}{Peter D{\"u}ben}, {and} \bibinfo{person}{Tim
  Palmer}.} \bibinfo{year}{2019}\natexlab{}.
\newblock \showarticletitle{Accelerating high-resolution weather models with
  deep-learning hardware}. In \bibinfo{booktitle}{{\em Proceedings of the
  Platform for Advanced Scientific Computing Conference}}.
  \bibinfo{pages}{1--11}.
\newblock


\bibitem[\protect\citeauthoryear{Huang, Yu, and van~de Geijn}{Huang
  et~al\mbox{.}}{2018}]%
        {Huang2018ImplementingSA}
\bibfield{author}{\bibinfo{person}{Jianyu Huang}, \bibinfo{person}{Chenhan~D.
  Yu}, {and} \bibinfo{person}{Robert~A. van~de Geijn}.}
  \bibinfo{year}{2018}\natexlab{}.
\newblock \showarticletitle{Implementing Strassen's Algorithm with CUTLASS on
  NVIDIA Volta GPUs}.
\newblock \bibinfo{journal}{{\em ArXiv\/}}  \bibinfo{volume}{abs/1808.07984}
  (\bibinfo{year}{2018}).
\newblock


\bibitem[\protect\citeauthoryear{Ioannou, Robertson, Cipolla, and
  Criminisi}{Ioannou et~al\mbox{.}}{2017}]%
        {ioannou2017deep}
\bibfield{author}{\bibinfo{person}{Yani Ioannou}, \bibinfo{person}{Duncan
  Robertson}, \bibinfo{person}{Roberto Cipolla}, {and} \bibinfo{person}{Antonio
  Criminisi}.} \bibinfo{year}{2017}\natexlab{}.
\newblock \showarticletitle{Deep roots: Improving cnn efficiency with
  hierarchical filter groups}. In \bibinfo{booktitle}{{\em Proceedings of the
  IEEE conference on computer vision and pattern recognition}}.
  \bibinfo{pages}{1231--1240}.
\newblock


\bibitem[\protect\citeauthoryear{Iserles}{Iserles}{1999}]%
        {Iserles99howlarge}
\bibfield{author}{\bibinfo{person}{Arieh Iserles}.}
  \bibinfo{year}{1999}\natexlab{}.
\newblock \bibinfo{title}{How Large is the Exponential of a Banded Matrix?}
\newblock   (\bibinfo{year}{1999}).
\newblock


\bibitem[\protect\citeauthoryear{Jaderberg, Vedaldi, and Zisserman}{Jaderberg
  et~al\mbox{.}}{2014}]%
        {jaderberg2014speeding}
\bibfield{author}{\bibinfo{person}{Max Jaderberg}, \bibinfo{person}{Andrea
  Vedaldi}, {and} \bibinfo{person}{Andrew Zisserman}.}
  \bibinfo{year}{2014}\natexlab{}.
\newblock \showarticletitle{Speeding up convolutional neural networks with low
  rank expansions}.
\newblock \bibinfo{journal}{{\em arXiv preprint arXiv:1405.3866\/}}
  (\bibinfo{year}{2014}).
\newblock


\bibitem[\protect\citeauthoryear{Kal{\'e} and Krishnan}{Kal{\'e} and
  Krishnan}{1993}]%
        {Kal1993CHARMAP}
\bibfield{author}{\bibinfo{person}{Laxmikant~V. Kal{\'e}} {and}
  \bibinfo{person}{Sanjeev Krishnan}.} \bibinfo{year}{1993}\natexlab{}.
\newblock \showarticletitle{CHARM++: a portable concurrent object oriented
  system based on C++}. In \bibinfo{booktitle}{{\em OOPSLA '93}}.
\newblock


\bibitem[\protect\citeauthoryear{Lakshminarayana and Kim}{Lakshminarayana and
  Kim}{2014}]%
        {lakshminarayana2014spare}
\bibfield{author}{\bibinfo{person}{Nagesh~B Lakshminarayana} {and}
  \bibinfo{person}{Hyesoon Kim}.} \bibinfo{year}{2014}\natexlab{}.
\newblock \showarticletitle{Spare register aware prefetching for graph
  algorithms on GPUs}. In \bibinfo{booktitle}{{\em 2014 IEEE 20th International
  Symposium on High Performance Computer Architecture (HPCA)}}. IEEE,
  \bibinfo{pages}{614--625}.
\newblock


\bibitem[\protect\citeauthoryear{Li, Liang, Yan, Jia, and Li}{Li
  et~al\mbox{.}}{2019}]%
        {li2019coordinated}
\bibfield{author}{\bibinfo{person}{Xiuhong Li}, \bibinfo{person}{Yun Liang},
  \bibinfo{person}{Shengen Yan}, \bibinfo{person}{Liancheng Jia}, {and}
  \bibinfo{person}{Yinghan Li}.} \bibinfo{year}{2019}\natexlab{}.
\newblock \showarticletitle{A coordinated tiling and batching framework for
  efficient GEMM on GPUs}. In \bibinfo{booktitle}{{\em Proceedings of the 24th
  Symposium on Principles and Practice of Parallel Programming}}.
  \bibinfo{pages}{229--241}.
\newblock


\bibitem[\protect\citeauthoryear{Lin, Reinhardt, and Burger}{Lin
  et~al\mbox{.}}{2001}]%
        {DBLP:conf/hpca/LinRB01}
\bibfield{author}{\bibinfo{person}{Wei{-}Fen Lin}, \bibinfo{person}{Steven~K.
  Reinhardt}, {and} \bibinfo{person}{Doug Burger}.}
  \bibinfo{year}{2001}\natexlab{}.
\newblock \showarticletitle{Reducing {DRAM} Latencies with an Integrated Memory
  Hierarchy Design}. In \bibinfo{booktitle}{{\em Proceedings of the Seventh
  International Symposium on High-Performance Computer Architecture (HPCA'01),
  Nuevo Leone, Mexico, January 20-24, 2001}}. \bibinfo{publisher}{{IEEE}
  Computer Society}, \bibinfo{pages}{301--312}.
\newblock
\showDOI{%
\url{https://doi.org/10.1109/HPCA.2001.903272}}


\bibitem[\protect\citeauthoryear{Markidis, Der~Chien, Laure, Peng, and
  Vetter}{Markidis et~al\mbox{.}}{2018}]%
        {markidis2018nvidia}
\bibfield{author}{\bibinfo{person}{Stefano Markidis},
  \bibinfo{person}{Steven~Wei Der~Chien}, \bibinfo{person}{Erwin Laure},
  \bibinfo{person}{Ivy~Bo Peng}, {and} \bibinfo{person}{Jeffrey~S Vetter}.}
  \bibinfo{year}{2018}\natexlab{}.
\newblock \showarticletitle{Nvidia tensor core programmability, performance \&
  precision}. In \bibinfo{booktitle}{{\em 2018 IEEE International Parallel and
  Distributed Processing Symposium Workshops (IPDPSW)}}. IEEE,
  \bibinfo{pages}{522--531}.
\newblock


\bibitem[\protect\citeauthoryear{Mukunoki and Ogita}{Mukunoki and
  Ogita}{2020}]%
        {MUKUNOKI2020112701}
\bibfield{author}{\bibinfo{person}{Daichi Mukunoki} {and}
  \bibinfo{person}{Takeshi Ogita}.} \bibinfo{year}{2020}\natexlab{}.
\newblock \showarticletitle{Performance and energy consumption of accurate and
  mixed-precision linear algebra kernels on GPUs}.
\newblock \bibinfo{journal}{{\it J. Comput. Appl. Math.}}
  \bibinfo{volume}{372} (\bibinfo{year}{2020}), \bibinfo{pages}{112701}.
\newblock
\showISSN{0377-0427}


\bibitem[\protect\citeauthoryear{Navarro, Carrasco, Barrientos, Riquelme, and
  Vega}{Navarro et~al\mbox{.}}{2020}]%
        {navarro2020gpu}
\bibfield{author}{\bibinfo{person}{Cristobal~A Navarro},
  \bibinfo{person}{Roberto Carrasco}, \bibinfo{person}{Ricardo~J Barrientos},
  \bibinfo{person}{Javier~A Riquelme}, {and} \bibinfo{person}{Raimundo Vega}.}
  \bibinfo{year}{2020}\natexlab{}.
\newblock \showarticletitle{GPU Tensor Cores for fast Arithmetic Reductions}.
\newblock \bibinfo{journal}{{\em arXiv: Distributed, Parallel, and Cluster
  Computing\/}} (\bibinfo{year}{2020}).
\newblock


\bibitem[\protect\citeauthoryear{Odashima, Prado, and Vernek}{Odashima
  et~al\mbox{.}}{2017}]%
        {ODASHIMA2017}
\bibfield{author}{\bibinfo{person}{Mariana~M. Odashima},
  \bibinfo{person}{Beatriz~G. Prado}, {and} \bibinfo{person}{E. Vernek}.}
  \bibinfo{year}{2017}\natexlab{}.
\newblock \showarticletitle{{Pedagogical introduction to equilibrium Green's
  functions: condensed-matter examples with numerical implementations}}.
\newblock \bibinfo{journal}{{\em {Revista Brasileira de Ensino de
  F\~A\-sica}\/}}  \bibinfo{volume}{39} (\bibinfo{date}{00}
  \bibinfo{year}{2017}).
\newblock
\showISSN{1806-1117}
\showURL{%
\url{http://www.scielo.br/scielo.php?script=sci_arttext&pid=S1806-11172017000100402&nrm=iso}}


\bibitem[\protect\citeauthoryear{Olivaresamaya, Watson, Edgar, Vogt, Shao, and
  Aspuruguzik}{Olivaresamaya et~al\mbox{.}}{2010}]%
        {olivaresamaya2010accelerating}
\bibfield{author}{\bibinfo{person}{Roberto Olivaresamaya},
  \bibinfo{person}{Mark~A Watson}, \bibinfo{person}{Richard~G Edgar},
  \bibinfo{person}{Leslie Vogt}, \bibinfo{person}{Yihan Shao}, {and}
  \bibinfo{person}{Alan Aspuruguzik}.} \bibinfo{year}{2010}\natexlab{}.
\newblock \showarticletitle{Accelerating Correlated Quantum Chemistry
  Calculations Using Graphical Processing Units and a Mixed Precision Matrix
  Multiplication Library}.
\newblock \bibinfo{journal}{{\em Journal of Chemical Theory and Computation\/}}
  \bibinfo{volume}{6}, \bibinfo{number}{1} (\bibinfo{year}{2010}),
  \bibinfo{pages}{135--144}.
\newblock


\bibitem[\protect\citeauthoryear{Rubensson}{Rubensson}{2017}]%
        {Rubensson2017ChunksAT}
\bibfield{author}{\bibinfo{person}{Emanuel~H. Rubensson}.}
  \bibinfo{year}{2017}\natexlab{}.
\newblock \showarticletitle{Chunks and Tasks}.
\newblock


\bibitem[\protect\citeauthoryear{Rudberg, Rubensson, Sa{\l}ek, and
  Kruchinina}{Rudberg et~al\mbox{.}}{2018}]%
        {rudberg2018ergo}
\bibfield{author}{\bibinfo{person}{Elias Rudberg}, \bibinfo{person}{Emanuel~H
  Rubensson}, \bibinfo{person}{Pawe{\l} Sa{\l}ek}, {and}
  \bibinfo{person}{Anastasia Kruchinina}.} \bibinfo{year}{2018}\natexlab{}.
\newblock \showarticletitle{Ergo: An open-source program for linear-scaling
  electronic structure calculations}.
\newblock \bibinfo{journal}{{\em SoftwareX\/}}  \bibinfo{volume}{7},
  \bibinfo{pages}{107--111}.
\newblock


\bibitem[\protect\citeauthoryear{Ryoo, Rodrigues, Baghsorkhi, Stone, Kirk, and
  Hwu}{Ryoo et~al\mbox{.}}{2008}]%
        {ryoo2008optimization}
\bibfield{author}{\bibinfo{person}{Shane Ryoo}, \bibinfo{person}{Christopher~I
  Rodrigues}, \bibinfo{person}{Sara~S Baghsorkhi}, \bibinfo{person}{Sam~S
  Stone}, \bibinfo{person}{David~B Kirk}, {and} \bibinfo{person}{Wenmei~W
  Hwu}.} \bibinfo{year}{2008}\natexlab{}.
\newblock \showarticletitle{Optimization principles and application performance
  evaluation of a multithreaded GPU using CUDA}. \bibinfo{pages}{73--82}.
\newblock


\bibitem[\protect\citeauthoryear{Schuch, Cirac, and Wolf}{Schuch
  et~al\mbox{.}}{2006}]%
        {schuch2006quantum}
\bibfield{author}{\bibinfo{person}{Norbert Schuch}, \bibinfo{person}{J~Ignacio
  Cirac}, {and} \bibinfo{person}{Michael~M Wolf}.}
  \bibinfo{year}{2006}\natexlab{}.
\newblock \showarticletitle{Quantum States on Harmonic Lattices}.
\newblock \bibinfo{journal}{{\em Communications in Mathematical Physics\/}}
  \bibinfo{volume}{267}, \bibinfo{number}{1} (\bibinfo{year}{2006}),
  \bibinfo{pages}{65--92}.
\newblock


\bibitem[\protect\citeauthoryear{Simon}{Simon}{1982}]%
        {simon1982some}
\bibfield{author}{\bibinfo{person}{Barry Simon}.}
  \bibinfo{year}{1982}\natexlab{}.
\newblock \showarticletitle{Some Jacobi matrices with decaying potential and
  dense point spectrum}.
\newblock \bibinfo{journal}{{\em Communications in Mathematical Physics\/}}
  \bibinfo{volume}{87}, \bibinfo{number}{2} (\bibinfo{year}{1982}),
  \bibinfo{pages}{253--258}.
\newblock


\bibitem[\protect\citeauthoryear{Tessera}{Tessera}{2010}]%
        {tessera2010left}
\bibfield{author}{\bibinfo{person}{Romain Tessera}.}
  \bibinfo{year}{2010}\natexlab{}.
\newblock \showarticletitle{Left inverses of matrices with polynomial decay}.
\newblock \bibinfo{journal}{{\em Journal of Functional Analysis\/}}
  \bibinfo{volume}{259}, \bibinfo{number}{11} (\bibinfo{year}{2010}),
  \bibinfo{pages}{2793--2813}.
\newblock


\bibitem[\protect\citeauthoryear{Van De~Geijn and Watts}{Van De~Geijn and
  Watts}{1997}]%
        {van1997summa}
\bibfield{author}{\bibinfo{person}{Robert~A Van De~Geijn} {and}
  \bibinfo{person}{Jerrell Watts}.} \bibinfo{year}{1997}\natexlab{}.
\newblock \showarticletitle{SUMMA: Scalable universal matrix multiplication
  algorithm}.
\newblock \bibinfo{journal}{{\em Concurrency: Practice and Experience\/}}
  \bibinfo{volume}{9}, \bibinfo{number}{4} (\bibinfo{year}{1997}),
  \bibinfo{pages}{255--274}.
\newblock


\bibitem[\protect\citeauthoryear{Volkov and Demmel}{Volkov and Demmel}{2008}]%
        {volkov2008benchmarking}
\bibfield{author}{\bibinfo{person}{Vasily Volkov} {and} \bibinfo{person}{James
  Demmel}.} \bibinfo{year}{2008}\natexlab{}.
\newblock \showarticletitle{Benchmarking GPUs to tune dense linear algebra}.
\newblock  (\bibinfo{year}{2008}), \bibinfo{pages}{31}.
\newblock


\bibitem[\protect\citeauthoryear{Williams}{Williams}{2012}]%
        {10.1145/2213977.2214056}
\bibfield{author}{\bibinfo{person}{Virginia~Vassilevska Williams}.}
  \bibinfo{year}{2012}\natexlab{}.
\newblock \showarticletitle{Multiplying Matrices Faster than
  Coppersmith-Winograd}. In \bibinfo{booktitle}{{\em Proceedings of the
  Forty-Fourth Annual ACM Symposium on Theory of Computing}} {\em
  (\bibinfo{series}{STOC ’12})}. \bibinfo{publisher}{Association for
  Computing Machinery}, \bibinfo{address}{New York, NY, USA},
  \bibinfo{pages}{887–898}.
\newblock
\showISBNx{9781450312455}
\showDOI{%
\url{https://doi.org/10.1145/2213977.2214056}}


\bibitem[\protect\citeauthoryear{Ye}{Ye}{2013}]%
        {ye2013error}
\bibfield{author}{\bibinfo{person}{Qiang Ye}.} \bibinfo{year}{2013}\natexlab{}.
\newblock \showarticletitle{Error Bounds for the Lanczos Methods for
  Approximating Matrix Exponentials}.
\newblock \bibinfo{journal}{{\it SIAM J. Numer. Anal.}} \bibinfo{volume}{51},
  \bibinfo{number}{1} (\bibinfo{year}{2013}), \bibinfo{pages}{68--87}.
\newblock


\end{thebibliography}

%
\end{document}